\documentclass[useAMS,usenatbib,fleqn]{mn2e}
\usepackage{graphicx}
\usepackage{amsmath}
\usepackage{booktabs}
\usepackage{multirow}
\usepackage{multicol}
\usepackage{tabularx}
\usepackage{mathrsfs}
\usepackage{bm}
\def\gsim{\;\lower4pt\hbox{${\buildrel\displaystyle >\over\sim}$}\;}
\def\lsim{\;\lower4pt\hbox{${\buildrel\displaystyle <\over\sim}$}\;}
\def\grls{\;\lower4pt\hbox{${\buildrel\displaystyle >\over <}$}\;}
\def\d{\mathrm{d}}
\def\vvec{\mathbf{v}}
\def\bvec{\mathbf{B}}
\def\evec{\mathbf{E}}
\def\curl{\nabla\times}
\def\div{\nabla\cdot}
\def\10t{10^}
\def\t10t{\times 10^}
\def\cps{\mathrm{\ cm\ s}^{-1}}
\def\rvec{\mathbf{r}}
\def\phivec{{\bm \phi}}
\def\thvec{{\bm \theta}}
\def\kps{\mathrm{\ km\ s}^{-1}}
\def\erg{\mathrm{\ erg}}
\def\cscor{\mathrm{Corr}_\mathrm{cs}}
\def\gauss{\mathrm{\ G}}
\title[Steady Axisymmetric 2.5D MHD]
  {Steady-State Axisymmetric MHD Solutions\\
   with Various Boundary Conditions}
\author[Lile Wang and Yu-Qing Lou]{
 Lile Wang$^{1}$
 and Yu-Qing Lou$^{1}$
 \thanks{E-mail: louyq@tsinghua.edu.cn (Y-QL)}
\\
$^1$Department of Physics and Tsinghua Centre for Astrophysics
 (THCA), Tsinghua University, Beijing 100084, China \\
}

\date{Accepted 2014 January 08. Received 2014 January 08;
    in original form 2013 July 14}
\pubyear{2012}

\begin{document}

\maketitle

\begin{abstract}
  Axisymmetric magnetohydrodynamics (MHD) can be invoked
    for describing astrophysical magnetized flows and
    formulated to model stellar magnetospheres including
    main sequence stars (e.g. the Sun), compact stellar
    objects [e.g. magnetic white dwarfs (MWDs), radio pulsars,
    anomalous X-ray pulsars (AXPs), magnetars, isolated
    neutron stars etc.], and planets as a major step forward
    towards a full three-dimensional model construction.
  Using powerful and reliable numerical solvers based on two
  distinct finite-difference method (FDM) and finite-element
  method (FEM) schemes of algorithm, we examine
  axisymmetric steady-state or stationary MHD models
    in \citet{2001GApFD..94..249T}, finding that their
    separable semi-analytic nonlinear solutions are actually
    not unique given their specific selection of several
    free functionals and chosen boundary conditions.
  Similar situations of multiple nonlinear solutions
    with the same boundary conditions actually
    also happen to force-free magnetic field
    models of \citet{1990ApJ...352..343L}.
  The multiplicity of nonlinear steady MHD solutions gives
    rise to differences in the total energies contained in
    the magnetic fields and flow velocity fields as well as
    in the asymptotic behaviours approaching infinity, which
    may in turn explain why numerical solvers tend to
    converge to a nonlinear solution with a lower energy
    than the corresponding separable semi-analytic one.
  By properly adjusting model parameters, we invoke
    semi-analytic and numerical solutions to describe
    different kind of scenarios, including nearly parallel
    case and the situation in which the misalignment between
    the fluid flow and magnetic field is considerable.
  We propose that these MHD models are capable of describing
    the magnetospheres of MWDs as examples of applications
    with moderate conditions (including magnetic field)
    where the typical values of several important
    parameters are consistent with observations.
  Physical parameters can also be estimated based
    on such MHD models directly.
  We discuss the challenges of developing numerical
    extrapolation MHD codes in view of nonlinear
    degeneracy.
%{\it The parameters can also be evaluated by starting from models
%  directly.}
%{\bf It would be better to discuss
%  and refer to this paper more fully in the main text in terms of
%  isolated neutron stars!}
%
%   {\bf Please check references of isolated neutron stars;
%    they tend to be slow rotator; about 10 or so detected so far
%    Marten Van Kerkwijk and Thompson work. mentioned Feb 19, 2013.
%    They should also have magnetospheres. }
\end{abstract}
\begin{keywords}
  magnetohydrodynamics (MHD) -- magnetic fields -- stars: neutron --
  white dwarfs -- stars: winds, outflows -- methods: numerical
% checked with the new list from Oxford University Press (OUP) Dec 16, 2013
% {\bf Please select up to six key words from MNRAS list!!}
\end{keywords}

\section{Introduction}
\label{sec:introduction}

Magnetohydrodynamics (MHD) is often invoked to
  describe dynamic motions of (fully or partially)
  ionized and magnetized plasmas in broad contexts
  of astrophysical systems.
Since hydrodynamic equations for such a magnetized
  plasma are coupled with the Maxwell equations of
  electrodynamics \citep[e.g.][]{1975clel.book.....J},
  direct solutions to this set of MHD nonlinear partial
  differential equations (PDEs) are extremely
  challenging and notoriously complicated, which are
  usually obtained through numerical simulations.
For favorable and special situations, when some
  additional conditions such as ideal MHD, steady
  state and spatial symmetries (especially the type
  of symmetries that can reduce the problem to a
  2.5-dimensional one, e.g. helical symmetry, translational
  symmetry or axisymmetry) are assumed, the system of MHD
  PDEs can be significantly simplified after tedious yet
  straightforward algebraic and calculus manipulations \citep[e.g.][]{1981ApJ...245..764T,1982ApJ...252..775T, 1982ApJ...259..820T}.
Vector nonlinear ideal MHD PDEs can be reduced to scalar
  ones by making use of available conservation laws and
  by considerable simplifications due to symmetries.
There have been various attempts to construct semi-analytic
  nonlinear solutions under allowed conditions.
The scenarios of those attempts range from static force-free
  cases \citep[e.g.][]{1990ApJ...352..343L} to MHD cases
  with dynamic flows \citep[e.g.][]{1986ApJ...302..163L,
  1988Ap&SS.149..217K,1989ApJ...342.1028T, 1994ApJ...429..139C,
  1996A&A...310..341D, 2004PhPl...11.3015K,2007arXiv0704.1758T}.

In this paper, we adopt the ideal MHD formulation and approach
  of \citet{1998PhPl....5.2378T} and especially of
  \citet{2001GApFD..94..249T}, in which axisymmetry and
  incompressible conditions are both imposed to further
  reduce the problem, while the gravitational field of
  a central stellar object and gas flows are retained
  for sensible physical reality.
The idealizations and semi-analytic results form the key
  basis of our further numerical explorations and elaborations.

Those MHD PDEs
%  (especially non-perturbative ones)
  are nonlinear, raising the possibility that their
  solutions may not be unique for a given set of
  specified boundary conditions in general.
In the current context, this specifically relates
  to the important issue of boundary conditions
  at large radii for the extrapolation of solar
  and stellar coronal magnetic field configurations.
By our independent numerical explorations using
  distinct finite-difference method (FDM) and
  finite-element method (FEM) schemes, we noticed that
  a class of separable semi-analytic solutions to the force-free
  magnetic field problem of \citet[][]{1990ApJ...352..343L} are
  indeed not unique with the same nonlinear PDEs and spatial
  boundary conditions, which in fact substantially extended
  and confirmed the results of \citet{1993A&A...278..589B}
  (see Appendix \ref{sec:prev-works-multi} for more details).
%  {\bf Please check and comment on the critical $n$
%  for stability of semi-analytical solutions in
%  Bruma \& Cuperman (1990)!}
%{\it
A specific index $n$ in \citet{1993A&A...278..589B}
  (see also Appendix \ref{sec:prev-works-multi} here)
  turns out to be critical for the presence of multiple
  nonlinear solutions with the same boundary conditions.
When this $n$ is greater than a critical value $n_c$, the
  solution is unique; when its value decreases such that
  $n<n_c$, the solution ``bifurcates'' into two branches,
  even though the boundary conditions for the two solution
  branches remain identical.
% }
Mathematically,
  the type of quasi-linear MHD differential equation
  in \citet{2001GApFD..94..249T} may not be excluded
  for possibly possessing multiple solutions when
  the functional corresponding to the original PDE
  is not convex \citep[see][for the pertinent
  functional analysis]{2011PDE..book.Taylor}.
%  \citep[see][]{2011PDE..book.Taylor}.
%  {\bf Explicit definition and clear meaning?}
%
These hints and clues motivate us to carefully examine
  the uniqueness of solutions by solving the nonlinear
  MHD PDEs numerically in reference to the available
  semi-analytical solutions.
Numerical solvers implemented with quite different algorithms
  [e.g., FDM and FEM] and variable grid resolutions are applied to
  perform necessary mutual cross-checks, which would give us
  reasonable confidence on our numerical findings that a
  definite type of different numerical solutions may exist
  with the same boundary conditions and nonlinear MHD PDEs
  as in \citet{2001GApFD..94..249T}.

Axisymmetric nonlinear ideal MHD PDEs are often invoked to
  describe a variety of astrophysical magnetized plasma
  systems, such as magnetized accretion discs, magnetized
  stellar winds, magnetized astrophysical jets on completely
  different scales
  \citep[e.g.][]{1986ApJS...62....1L,1986ApJ...302..163L,
      1988Ap&SS.149..217K,1989ApJ...342.1028T,
      1994ApJ...429..139C,2003ApJ...599L...5P},
  and magnetospheres of stars (e.g., the Sun), planets
  (e.g., Earth, Jupiter, Saturn etc.) and compact
  stellar objects (e.g., radio pulsars, magnetars, AXPs,
  isolated neutron stars, cataclysmic variables etc.)
  \citep{1973ApJ...180L.133M,1999ApJ...511..351C, 2006MNRAS.367...19K}.
Compact stellar objects, including white dwarfs and
  neutron stars etc., possess intense magnetic fields
  as inferred by extensive observations.
% that are excessively intense.
For neutron stars (e.g. radio pulsars, AXPs, magnetars,
  isolated neuron stars), whose magnetic field intensities
  may reach as high as $10^{11}\gauss \sim 10^{15}\gauss$,
%  {\bf How about magnetars of up to $\sim 10^{15}\gauss$?}
  the so-called force-free conditions are often presumed
  since the magnetic Lorentz force can be hardly balanced
  by other known forces (such as gravity and pressure force).
Meanwhile,
  velocity of magnetized fluid particles are usually so
  large that relativistic conditions could also apply
  \citep[e.g.][]{1992ApJ...397L..67L, 1993ApJ...414..656L,
    1993ApJ...418..709L, 1998MNRAS.294..443L, 2006ApJ...648L..51S}.
%WLL   {\bf Be careful with pertinent contexts.}
However, for magnetic white dwarfs, the intensity of magnetic field
  is significantly lower around $\sim 10^4 - 10^9\gauss$ \citep[e.g.] []
  {1995ApJ...448..305S}, in which we would expect that the force-free
  condition may not hold and thus the misalignment between the fluid
  flow and magnetic field may possibly be considerable.
%WLL  {\bf What is the point? Be careful with the wording!}
We expect that, by properly selecting parameters, the
  semi-analytic MHD models of \citet{2001GApFD..94..249T} and
  our corresponding numerical models may offer reasonable
  though idealized descriptions for magnetospheres of magnetic
  white dwarfs (MWDs) in general.

%{\bf Stop here!}

This paper is structured as follows.
 Section \ref{sec:formulation} presents the basic
  model formulation of the simplified and reduced
  ideal MHD problem.
 Key steps in derivations and reductions of MHD PDEs
  to a single scalar PDE involving free functionals
  are shown in Section \ref{sec:reduction-mhd}.
In Section \ref{sec:numerical-results}, semi-analytic
  solutions and their corresponding numerical model
  solutions are presented to compare with the pertinent
  computational results.
Section \ref{sec:applicatoins} offers analyses and
  astrophysical applications of our ideal MHD models.
We summarize results and conclusions in Section
  \ref{sec:conclusions-summary}.
Mathematical derivations are detailed in
  three Appendices for the convenience of reference.

%{\bf Stop here!}

\section{Basic MHD Model Formulation}
\label{sec:formulation}

Steady-state (i.e. stationary or time-independent) ideal
  MHD is governed by the following set of nonlinear PDEs
  in c.g.s.-Gaussian units, which combine the Maxwell
  equations and the Eulerian hydrodynamic equations with
  certain approximations \citep[e.g.][]{1981ApJ...245..764T}:
\begin{equation}
  \label{eq:mhd-equ}
  \begin{split}
    & \div (\rho \vvec) = 0\ ,
    \\
    & \rho (\vvec \cdot \nabla) \vvec = \dfrac{(\curl\bvec )
    \times \bvec}{4\pi} - \nabla P - \rho \dfrac{G
      \mathcal{M}}{r^2}\mathbf{\hat r} \ ,
    \\
    & \evec + \dfrac{\vvec \times \bvec}{c} = 0\ ,
    \qquad\curl \evec = 0\ ,\qquad \div \bvec = 0\ ,
  \end{split}
  \end{equation}
where $\hat\rvec$ is the unit vector along the radial direction
  pointing outwards, $r$ is the radius, $\rho$ is the fluid mass
  density, $\mathcal{M}$ is the central stellar mass, $P$ is the
  plasma pressure, $\vvec$ is the bulk flow velocity vector,
  $\evec$ and $\bvec$ are the electric and magnetic vector fields
  respectively, $c= 3\t10t{10}\cps$ is the speed of light, and
  $G=6.67\t10t{-8} \mathrm{\ dyn\ cm}^{2}\mathrm{\ g}^{-2}$ is
  the universal gravitational constant.
The ideal MHD approximation
% without the self-gravity
  is adopted here to ignore effects of resistivity, viscosity,
  thermal conduction and radiative losses and so forth, while
  the gravitational field of a central spherical stellar object
  (e.g. a main-sequence star, a neutron star or a white dwarf)
  of mass $\mathcal{M}$ is included.
We shall not go into the stellar interior and thus regard
  the spherical stellar surface as the inner boundary
  denoted by a reference radius $r_0$.
The change of shape for the stellar surface due to rotation
  is ignored in the present treatment.
In this formalism, we analyze pertient problems in spherical
  polar coordinates $(r,\ \theta,\ \phi)$ for radius $r$,
  polar angle $\theta$ and azimuthal angle $\phi$, respectively
  (n.b. the notations $\Phi$ and $\varphi$ introduced later
  bear different meanings than $\phi$ in our formalism).
We use $\mu\equiv\cos\theta$ to denote the cosine of
  $\theta$ for the convenience of mathematical analysis.

We note a few basic invariance properties of
  stationary MHD PDE system \eqref{eq:mhd-equ}
  under several transformations.
First, for a simultaneous transformation of $\vvec$
  to $-\vvec$ and $\evec$ to $-\evec$ with other
  variables unchanged, MHD PDE system
  \eqref{eq:mhd-equ} remains invariant.
Secondly, for a simultaneous transformation of $\bvec$
  to $-\bvec$ and $\evec$ to $-\evec$ with other
  variables unchanged, nonlinear MHD PDE system
  \eqref{eq:mhd-equ} remains invariant.
Thirdly, for a simultaneous transformation of
  $\bvec$ to $-\bvec$ and $\vvec$ to $-\vvec$
  with other variables unchanged, MHD PDE system
  \eqref{eq:mhd-equ} remains invariant.
These properties can be readily understood on
  intuitive physical terms for ideal MHD.

\section{Derivation for steady ideal MHD PDEs
  with axisymmetry and incompressible fluid }
%  approximation}
\label{sec:reduction-mhd}

The formalism in the foregoing section
 is sufficiently general.
When axisymmetry and incompressibility are imposed,
  system \eqref{eq:mhd-equ} of nonlinear ideal MHD
  PDEs can be considerably simplified.
In this section, we briefly review the simplification
  steps in \citet{1982ApJ...252..775T,1998PhPl....5.2378T}
  and especially \citet{2001GApFD..94..249T} (n.b. we
  have properly recast some pertinent expressions in
  c.g.s.-Gaussian units in a consistent manner).
Details of derivations and reductions relevant
  to this section are included in Appendix
  \ref{sec:scalar-pde-derivation} for reference.

The first integrals (or equivalently, several conservation
  laws due to the axisymmetry) of nonlinear ideal MHD PDEs
  \eqref{eq:mhd-equ} can be then derived.
The two vector fields
  $\bvec$ (magnetic field) and $\rho\vvec$ (mass flux
  density) can be expressed in the following forms by
  introducing four free scalar functions of $r$ and
  $\theta$, namely $\psi$, $I$, $\Theta$ and $F$
  \citep[see][here $\hat\phivec$ is the unit vector along
  the azimuthal $\phi$ direction for spherical polar
  coordinates] {1981ApJ...245..764T, 1998PhPl....5.2378T}:
\begin{equation}
  \label{eq:vect-quant}
  \begin{split}
    & \evec = -\nabla \Phi\ ,
    \\
    & \bvec = (4\pi)^{1/2} \left(
    \dfrac{I\ \hat\phivec}{r \sin\theta}
     + \dfrac{\hat\phivec
    \times \nabla \psi}{r\sin\theta} \right) \ ,
    \\
    & \rho \vvec = \dfrac{\Theta\ \hat\phivec}{r \sin\theta}
   + \dfrac{\hat\phivec \times \nabla F}{r\sin\theta}\ ,
  \end{split}
\end{equation}
% {\bf It would be better to prepare an appendix
% to describe this analysis in details!!}
where $\Phi$ is the steady electric potential, $\psi$
 is in the physical dimension of magnetic flux, $F$ is
 in the physical dimension of mass flux, $I$ is related
 to the toroidal
% (azimuthal) component of the
 magnetic field, and $\Theta$ is related to the toroidal
 (azimuthal) component of the mass flux density (or flow
 velocity) field.

Several other integrals of nonlinear MHD PDEs \eqref{eq:mhd-equ}
  require free function $F$ and the electric potential
  $\Phi$ to be full functionals of $\psi$.
For an incompressible fluid with $\div\vvec = 0$, we have
  $\vvec\cdot\nabla\rho = 0$ indicating that the mass
  density $\rho$ is yet another full functional of $\psi$
  \citep[see section 2 of][]{2001GApFD..94..249T}.
It then follows that constraints on those reduced
  scalar function fields are
\begin{equation}
  \label{eq:scalar-relation}
  \begin{split}
    & P_s(\psi) = P + \rho \left\{ \dfrac{v^2}{2}
      - \dfrac{G\mathcal{M}}{r} -
      \dfrac{r^2\sin^2\theta[c\Phi'(\psi)]^2}
      {4\pi(1-M^2)}
    \right\}\ ,
    \\
    & I F'(\psi) - \Theta =
    \dfrac{c\Phi'(\psi)}{(4\pi)^{1/2}}\rho(\psi) r^2
    \sin^2\theta\ ,
    \\
    & I \left[ 1 - \dfrac{F'(\psi)^2}{\rho} \right] + r^2
    \sin^2\theta F'(\psi) \dfrac{c\Phi'(\psi)}{(4\pi)^{1/2}}
    \equiv X(\psi)\
  \end{split}
\end{equation}
\citep[][] {2001GApFD..94..249T}.
Here the prime [such as $F'(\psi)$ as an example] indicates
  the first derivative of a free full functional with respect
  to $\psi$, $X(\psi)$ is another full functional of $\psi$
  involving the toroidal component of the magnetic field,
  the poloidal component of the mass flux density and the
  poloidal component of electric field, and $P_s(\psi)$ (as
  yet another full functional of $\psi$) stands for the total
  pressure involving the components of steady pressure,
  kinetic energy density, gravitational potential energy
  density and electromagnetic energy density.
The quantity $M$ is the poloidal Alfv\'enic Mach number
  to be defined presently.
The Alfv\'en velocity $v_\mathrm{Ap}$ associated with the
  poloidal magnetic field is defined by\footnote{
%WLL  {\bf They use cylindrical coordinates.
%  Still a typo? Or $4\pi$?}
There is
  a typo in \citet{2001GApFD..94..249T} for the expression
  of $v_\mathrm{Ap}^2$ for the square of the Alfv\'en
    velocity associated with the poloidal magnetic field.}
    $v_\mathrm{Ap}^2 = |\nabla\psi|^2/(\rho r^2\sin^2\theta)$.
Meanwhile, the magnitude squared of the poloidal component
  of plasma flow velocity $v_\mathrm{p}$ is given by
  $v_\mathrm{p}^2=|F'\nabla\psi|^2/(\rho\ r\sin\theta)^2$.
Together, they naturally give the defining expression
  for the poloidal Alfv\'enic Mach number squared as
  $M^2\equiv v_\mathrm{p}^2/v_\mathrm{Ap}^2=(F')^2/\rho$
  \citep[][]{1998PhPl....5.2378T}.
This mathematical formulation though in different forms
  should be equivalent to the standard Grad-Shafranov
  formulation.
In fact, it was referred to as Grad-Schl\"uter-Shafranov
  equation in Troumoulopoulos \& Tasso (2001) and as
  generalized Grad-Shafranov equation in Troumoulopoulos
  et al. (2007).

We now introduce an expedient transformation converting
  the magnetic flux function $\psi$ to function $U$ by
  an integral and therefore regard all free functionals
  of $\psi$ as functionals of $U$.
Specifically, the integral relation between
  $U$ and $\psi$ is simply given below by
\begin{equation}
  \label{eq:psi-u-transformation}
  U(\psi)=\int_0^\psi\left[1 - M^2(\tilde{\psi})
  \right]^{1/2}\d\tilde{\psi}\ .
\end{equation}
This integral transformation is only valid
  for $M^2<1$ (viz., sub-Alfv\'enic poloidal flow by
  the foregoing definition of $M$), and our subsequent
  analyses will be carried out in this designated
  regime of a sub-Alfv\'enic poloidal flow.
Such type of sub-Alfv\'enic poloidal MHD flows avoids
  any singularities and MHD critical surfaces.
In dealing with MHD flows passing across critical points,
  transformation (\ref{eq:psi-u-transformation}) should
  not be imposed globally and one needs to solve for
  $\psi$ directly.

After straightforward yet tedious algebraic
  and calculus manipulations, we arrive at a
  reduced elliptic PDE in terms of $U$ \citep[see]
  [n.b. the conversion between different systems of
  units adopted]{2001GApFD..94..249T}:
\begin{equation}
  \label{eq:u-pde-dimensional}
  \begin{split}
    0 & = \dfrac{\partial^2 U}{\partial r^2} + \left(
      \dfrac{1-\mu^2}{r^2} \right)
    \dfrac{\partial^2U}{\partial \mu^2} +
    \dfrac{1}{2}\dfrac{\d}{\d U}
    \left(\dfrac{X^2}{1-M^2}\right)
    \\
    &\qquad + r^2(1-\mu^2) \left( \dfrac{\d P_s}{\d U} +
      \dfrac{G\mathcal{M}}{r}\dfrac{\d \rho}{\d U} \right)
    \\
    &\qquad\qquad + r^4 (1-\mu^2)^2 \left( \dfrac{c^2}{4\pi} \right)
    \dfrac{\d}{\d U} \left[ \rho \left( \dfrac{\d \Phi}
    {\d U} \right)^2 \right]\ ,
  \end{split}
\end{equation}
where $\mu\equiv\cos\theta$.

%{\bf In this formalism, can we set $\vvec=0$ to study
%  magnetostatic solutions still involving gas pressure and
%  gravity.  For example, $\Phi=0$, $\Theta=0$, and $F=0$.
%  The last two lines in equation (3) are not necessary
%  and integral transformation (4) not necessary.  This
%  investigation should be very important. }

%{\it
We emphasize here that our formal reduction is also valid
  for describing the static case of $\vvec=0$, where we
  require more vanishing scalar fields: $\Phi$ (coming
  from $\evec=0$), $\Theta$, and $F$ (thus $M$).
As results of the absence of these terms,
  equation \eqref{eq:u-pde-dimensional} becomes
\begin{equation}
  \label{eq:u-pde-dimensional-v-vanish}
  \begin{split}
    0 & = \dfrac{\partial^2 U}{\partial r^2} +
      \left(\dfrac{1-\mu^2}{r^2} \right)
    \dfrac{\partial^2U}{\partial \mu^2} +
    \dfrac{1}{2}\dfrac{\d X^2}{\d U}
    \\ \qquad
    & \qquad\qquad + r^2(1-\mu^2) \left( \dfrac{\d P_s}{\d U} +
      \dfrac{G\mathcal{M}}{r}\dfrac{\d \rho}{\d U} \right)\ ,
  \end{split}
\end{equation}
where $U=\psi$ for $M=0$.
While some equations may appear singular with the
  static requirement of $\vvec=0$ (e.g. in equations
  \eqref{eq:v-field-component}), we note that this would not
  cause real contradictions, as can be verified by inserting
  $F=0$, $\Theta=0$ and $\Phi=0$ into pertinent equations.
Clearly, the last two lines in equation \eqref{eq:scalar-relation}
  are not necessary and integral transformation
  \eqref{eq:psi-u-transformation} is not needed.
This reduced static version \eqref{eq:u-pde-dimensional-v-vanish}
  would be important and useful in describing various possible
  magnetostatic equilibria with the gravity of the central
  stellar object, the plasma pressure, the magnetic pressure
  and the magnetic tension.
Pertinent astrophysical applications will be further
  explored and elaborated in future works.
%}

\subsection{Separable solutions of reduced MHD PDE}
\label{sec:separable-sol}

%In general,
Equation \eqref{eq:u-pde-dimensional} is an elliptic
  PDE, whose nonlinearity poses considerable challenges
  for finding separable solutions in conventional manners.
Nevertheless, there exist still some elegant procedures
  \citep[e.g.][]{2001GApFD..94..249T} to construct
  separable solutions by explicitly specifying the
  chosen forms of pertinent free functionals of $U$.
We present below one class of such
  separable nonlinear solutions.

We first recast MHD PDE \eqref{eq:u-pde-dimensional}
  into the following dimensionless form of
\begin{equation}
  \label{eq:u-pde-dimensionless}
  \begin{split}
    0 & = \eta^2\dfrac{\partial}{\partial \eta}\left( \eta^2
      \dfrac{\partial u}{\partial \eta} \right) +
    \eta^2(1-\mu^2) \dfrac{\partial^2u}{\partial \mu^2} +
    \dfrac{\alpha^2 \gamma_0}{2} \dfrac{\d}{\d u} (x^2)
    \\
    &\qquad + \dfrac{\alpha^2 \beta_0}{2} \left(
      \dfrac{1-\mu^2}{\eta^2} \right) \dfrac{\d p_s}{\d u} +
    \dfrac{\alpha^2 \epsilon_0} {2} \left(
      \dfrac{1-\mu^2}{\eta}\right) \dfrac{\d \varrho}{\d u}
    \\
    &\qquad\qquad + \alpha^2 \delta_0 \left( \dfrac{1-\mu^2}
      {\eta^2}\right)^2 \dfrac{\d}{\d u}\left[ \varrho
      \left( \dfrac{\d \varphi}{\d u}\right)^2 \right]\ ,
  \end{split}
\end{equation}
where $\alpha$ is a constant dimensionless parameter.
%WLL {\bf Explicit definition of $\alpha$? It appears
% that $\alpha^2$ may be systematically absorbed in
% the pertinent dimensionless parameters of the
% above equation?}
%{\it
In principle, this $\alpha$ parameter
  can be absorbed into other
  constants (viz. $\beta_0$, $\gamma_0$, $\epsilon_0$
  and $\delta_0$), but we keep it alone for the
  convenience of further discussions below [see
  equation \eqref{eq:u-separable-form}].
%WLL }
%  {\bf I am still curious for the reason of doing so.
%  June 30, 2013.}
In dimensionless reduced nonlinear PDE
  \eqref{eq:u-pde-dimensionless},
  we have actually made the following
  variables dimensionless in the forms of
% (here $r_0$, $U_0$, $X_0$, $P_{s0}$, $\rho_0$ and $\Phi_0$
% are fiducial constant dimensional physical parameters),
\begin{equation}
  \label{eq:dimensionless-trans}
  \begin{split}
    & \eta = \dfrac{r_0}{r}\ ,\qquad u = \dfrac{U}{U_0}\ ,\qquad
    x^2 = \dfrac{1}{(1-M^2)}\left(\dfrac{X}{X_0}\right)^2 \ ,
    \\
    & p_s = \dfrac{P_s}{P_{s0}}\ ,\ \qquad \varrho =
    \dfrac{\rho}{\rho_0}\ ,\ \qquad\varphi =
    \dfrac{\Phi}{\Phi_0}\ ,
  \end{split}
\end{equation}
  where $r_0$, $U_0$, $X_0$, $P_{s0}$, $\rho_0$ and $\Phi_0$
  are fiducial constant dimensional physical parameters.
  We also define the following four dimensionless parameters
  as [n.b. we adopt the Gaussian unit system; here
  $B_0 = \alpha (4\pi)^{1/2}U_0/r_0^2$ can be recognized
  as the fiducial value for the magnetic field strength]
\begin{equation}
  \label{eq:dimensionless-notes}
  \begin{split}
    & \epsilon_0 = \dfrac{G\mathcal{M}\rho_0}{r_0}
    \left(\dfrac{B_0^2}{8\pi} \right)^{-1}\!\!\! ,
    \ \qquad
    \gamma_0 = \dfrac{1}{2} \left( \dfrac{X_0}{r_0} \right)^2
    \left(\dfrac{B_0^2}{8\pi} \right)^{-1}\!\!\! ,
    \\
    & \delta_0 = \dfrac{\rho_0 c^2}{2} \left( \dfrac{\alpha
        \Phi_0}{B_0 r_0} \right)^2
        \left(\dfrac{B_0^2}{8\pi} \right)^{-1}\!\!\! ,\
    \quad  \beta_0 = P_{s0} \left(\dfrac{B_0^2}{8\pi}
    \right)^{-1}\!\!\! ,
  \end{split}
\end{equation}
  so that all variables in scalar nonlinear PDE
  \eqref{eq:u-pde-dimensionless} become dimensionless.
%  {\bf Need to check specifically! Physical
%  meanings of these dimensionless parameters.}
The physical meanings of the four dimensionless
    parameters defined in expression
    \eqref{eq:dimensionless-notes} are fairly clear;
they indicate the typical values of several important
    ratios by comparing pertinent physical quantities in
    pressure dimensions (e.g. pressure or energy density)
    with the magnetic pressure $P_B=B_0^2/(8\pi)$.
%The following three, in particular, have
They carry physical meanings for their
    numerators in the dimension of pressure:
    \begin{itemize}
    \item $\beta_0$: The total pressure $P_{s0}$
     versus the magnetic pressure (similar to the
     typical plasma $\beta$ parameter);
    \item $\delta_0$: Half of the mass-energy
     density versus the magnetic pressure multiplied
     by the energy density of the electric
      field versus the magnetic pressure;
    \item $\epsilon_0$: The gravitational energy density
     versus the magnetic pressure;
    \item $\gamma_0$: The functional $X(\psi)$ involves
     the toroidal component of the magnetic field and
     the interaction of the poloidal electric field
     and the poloidal mass flux density.
     We refer to $\gamma_0$ as the interaction
     energy density versus the magnetic energy density.
    \end{itemize}
%   {\bf How about $\gamma_0$?}

\citet{2001GApFD..94..249T} prescribed the functional
  forms of $x$, $p_s$, $\varrho$ and $\varphi$ as
  chosen powers of $u$ in order to construct separable
  semi-analytic solutions.
Here we slightly generalize their ansatz by taking the
  absolute value of $u$ (i.e. $|u|$) before entering the
  corresponding power functions, namely
\begin{equation}
  \label{eq:power-func-u}
  \begin{split}
    & x^2 = |u|^{2+2/\alpha}\ ,\qquad\qquad p_s = |u|^{2+4/\alpha}\ ,
    \\
    & \varrho = |u|^{2+3/\alpha}\ ,\qquad\qquad \varphi =
    \dfrac{2\alpha}{(2\alpha+3)} |u|^{1+3/(2\alpha)}\ .
  \end{split}
\end{equation}
%{\bf What is the reason of making such a choice?}
We particularly emphasize that the chosen forms of $p_s$ and
  $\varrho$ here actually represent a polytropic equation of
  state (EoS), i.e., $p_s$ is proportional to a certain power
  $\gamma$ of $\varrho$ and this power index
  $\gamma = 1 + 1/(2\alpha+3)$ is greater than
  unity for a positive $\alpha$ value (actually
  for as long as $\alpha>-3/2$).
This generalization of taking the absolute value
  of $u$ still enables the variable separation for
  nonlinear PDE \eqref{eq:u-pde-dimensionless}.
%{\it
Please note that the free functional $F(\psi)$ [or
  equivalently $F(u)$] is not prescribed here; the form of
  $F$ does not actually affect the solution of nonlinear
  PDE \eqref{eq:u-pde-dimensionless} at all.
However, $F$ is needed in recovering the axisymmetric
  vector field from the dimensionless solution as
  discussed in Appendix \ref{sec:field-recover}.
%}

For a positive $\alpha$ parameter, the derivatives of $x$,
  $p_s$, $\varrho$ and $\varphi$ with respect to $u$
%  {\bf absolute value?}
%  {\it (without the absolute value; what I want
%    to emphasize is that, generally, $\d |u|^\alpha/\d u$ might
%    have a jump point or diverge at  $u\rightarrow 0^\pm$)}
%    {\bf Need to check again!}
  are well defined as $u\rightarrow 0$.
Substituting functional form \eqref{eq:power-func-u} into
  dimensionless MHD PDE \eqref{eq:u-pde-dimensionless}
  and assuming $u$ in the following separable form of
\begin{equation}
  \label{eq:u-separable-form}
  u = h(\mu) \eta^\alpha\ ,
\end{equation}
  we can readily obtain a nonlinear ordinary differential equation
  (ODE) for $h(\mu)$ in the following form of\footnote{There exists
  a typo in \citet{2001GApFD..94..249T}, where the numerical
  factor 2 in front of the term with $\delta_0$ is somehow
  missed in their pertinent differential equation (27).}
\begin{equation}
  \label{eq:u-h-separated-ode}
  \begin{split}
    \dfrac{1}{h} \left( \dfrac{\d^2h}{\d \mu^2} \right)
    & = - \dfrac{\alpha (1+\alpha)}{(1-\mu ^2)}
    \left( 1 + \gamma_0 |h|^{2/\alpha }\right)
    \\
    & - \alpha (2+\alpha ) \beta _0 |h|^{4/\alpha } - \alpha
    \left(\dfrac{3}{2}+\alpha \right) \epsilon_0 |h|^{3/\alpha}\
    \\
    & - 2 \alpha (3+\alpha ) \left(1-\mu ^2\right) \delta_0
    |h|^{6/\alpha}\ .
  \end{split}
\end{equation}
%{\bf Need to check specifically}
%{\bf Absolute values on the LHS!}
This nonlinear ODE for $h(\mu)$ is invariant under
  the parity operation of $\mu\rightarrow -\mu$
  with respect to the equatorial plane.
Here, $\alpha$ parameter serves as the eigenvalue
  such that $h(\mu)$ satisfies the boundary
  conditions $h|_{\mu=\pm 1} = 0$ in order
  to avoid singularities at the two poles.
%  {\bf Direct tangible physical implications?}
Physically, $\alpha$ parameter is related to the
  polytropic index $\gamma$ in the polytropic EoS
  $p_s=\varrho^\gamma$ by $\gamma =1+1/(2\alpha+3)$.
%  {\bf Is the polytropic EoS actually
%  invoked in this formalism?}
The larger the $\alpha$ value, the closer the
 $\gamma$ value to unity.
We also note that, in contrast to \citet{1990ApJ...352..343L}
  for the problem of nonlinear force-free magnetic fields,
  here $h(\mu)$ cannot be rescaled under a different
  normalization
%  {\bf Reasons and explanations?}
  because there are several terms with different
  powers of $|h(\mu)|$ in summation.
Therefore, the values of first derivatives
  $h_{\mu 0}=\d h/\d \mu|_{\mu=\pm 1}$ at
  the two poles represent available degrees
  of freedom in our present MHD formulation.
This freedom can generate a rich nontrivial class of nonlinear
  MHD solutions with similar nodal structures but with different
  eigenvalues in a continuous manner.

%{\bf Please discuss the physical implications
%  of such plethora of MHD solutions with similar
%  nodal structures in $\mu$!}
%  {\bf What is the physical meaning of
%  $h_{\mu 0}=\d h/\d \mu|_{\mu=\pm 1}$
%  at the two poles? }
%  {\bf What are the conceivable physical
%  consequences for solar/stellar coronal
%  magnetic field configurations, stellar
%  magnetospheres, as well as (axisymmetric)
%  disk magnetic field geometries of such
%  continuous variations of
%  $h_{\mu 0}=\d h/\d \mu|_{\mu=\pm 1}$? }

%{\it
We comment that there exists a continuum of nonlinear
  eigensolutions, which feature the same topological
  structure (viz. having the same number of ``nodes''
  where $h=0$ in the interval $-1\leq\mu\leq 1$ ), by assigning
  a continuum of values for $\d h/\d \mu|_{\mu=\pm 1}$.
Those solutions have different $\alpha$ values (and thus
  different polytropic indices), while different $\d
  h/\d \mu$ are closely related to different magnitudes of
  radial components of $\bvec$ and $\vvec$ fields near the
  polar regions.
In other words, the configuration of fields of this type
  of solutions can vary continuously when the topological
  features of fields (e.g. dipolar fields, quadrupolar
  fields, etc.) remain the same.
This continuous variation is mathematically determined
  by and physically related to the magnitudes of radial
  components of magnetic and flow fields around $\mu=\pm 1$.
%}

%  {\bf Need a confirmation of this added statement.}
%{\bf Stop here!}

\section{Numerical Computational Results}
\label{sec:numerical-results}

\subsection{Descriptions of the numerical schemes}
\label{sec:num-scheme}

In practice, we numerically integrate nonlinear
  ODE \eqref{eq:u-h-separated-ode} for obtaining
  the angular variation $h(\mu)$
  by the well-known shooting scheme based on the
  standard fourth-order Runge-Kutta method
  \citep[e.g.][]{2002nrc..book.....P} with the
  specified boundary conditions at the two poles
  $\mu=\pm 1$.
%  {\bf Higher order available, e.g.  fifth-order
%  or sixth-order Runge-Kutta scheme?}
Higher-order Runge-Kutta numerical schemes are also available
  \citep[e.g.][]{Iserles200812}, but the fourth-order scheme
  adopted here is proven to be sufficiently
  accurate for our purpose of model/solution analysis.
We then combine the numerical solution of $h(\mu)$
  thus determined into separable solution form
  \eqref{eq:u-separable-form} to construct
  semi-analytic discrete nonlinear solutions of
  dimensionless PDE \eqref{eq:u-pde-dimensionless}
  for the prescribed free functionals.
This procedure is fairly straightforward to implement
  by concrete steps.

Meanwhile, we also attempt to directly solve two-dimensional
  dimensionless nonlinear PDE \eqref{eq:u-pde-dimensionless}
  by using two different numerical PDE solvers.
First, we build our own numerical code for solving
  quasilinear PDE using the FDM, that is, the relaxation
  method \citep[e.g.][section 20.5.2]{2002nrc..book.....P}
  implemented in the programming language C++.
Boundary conditions are specified at $\mu=\pm 1$ for the
  two poles, $\eta_i=r_0/r_i$ and $\eta_o=r_0/r_o$ where
  the italic subscripts ``$i$'' and ``$o$'' attached to
  the radial coordinate $r$ imply the ``inner'' and
  ``outer'' radii, respectively.
In other words, the numerical integration domain where the
  FDM PDE solver runs is $-1<\mu<1$ and $\eta_o<\eta<\eta_i$
  (n.b. here $r_o>r_i$ and hence $\eta_o<\eta_i$).
In order to avoid non-regular singularities, we
  simply set $u|_{\mu=\pm 1}=0$ at the two poles.
The inner boundary $\eta_i$ is generally set at $\eta_i=1$
  so that $r_0$ is simply the radius of the central
  spherical stellar object (e.g. a main-sequence
  star, or a neutron star or a white dwarf), while the
  outer boundary $\eta_o$ is generally set at
  $\eta_o\rightarrow 0^+$ corresponding to
  $r_o\rightarrow +\infty$ by intention
  (at least sufficiently large).
Our FDM numerical code involves $100$ grid points along
  the $\mu$ direction and $50$ grid points along the
  $\eta$ direction; these grid points are evenly spaced
  in the two spatial dimensions.
%WLL     {\bf Need to be precise here!}
%  {\bf Can you comment on the accuracy of the FDM code?
%  Please include the pertinent info. Email sent
%  and Skype conversation. Feb 19, 2013}
The accuracy of the FDM results are monitored in
  a real-time manner; the total error is estimated
  by summing the square of residuals on each grid
  point for every numerical iteration.
Each time of running the FDM solver, we take as the
  convergence criterion in that the average of
  absolute value of residuals of function value
  on each grid point is $\lsim 10^{-14}$ for
  relative differences.
For the purpose of testing and checking, we have also
  increased grid resolutions for the FDM solver and
  the resulting numerical solutions remain the same
  and stable.

%{\bf Stop here!}

Independently, we apply an open-source solver using the
  FEM, named \verb|FreeFem++| \citep{FreeFem++_Website},
  to verify the validity of the numerical solutions
  obtained by our own FDM solver code.
%{\it
This piece of software needs specific configuration files
  for initial input, which requires the user to give a
  well-defined functional zero-point problem in its
  pertinent syntax in a very specific manner.
We sketch the basic schemes for this configuration files,
  especially the mathematical essentials in Appendix
  \ref{sec:functioncal-ana-fem}, and refer the
  interested reader to the manual of \verb|FreeFem++|
  for detailed descriptions of the syntax.
%  }
In our numerical code construction, we assign
  $\sim 10^4$ elements which are automatically
  located by the FEM solver.
For the same purpose of testing and checking, we
  have also increased the number of elements for
  the FEM solver and the resulting numerical
  solutions remain the same and stable.
Although the two numerical methods of FDM and FEM are
  fundamentally different and independently implemented,
  the numerical solution results derived for the same
  problems agree with each other respectively in an
  almost perfect manner.

\subsection{Semi-analytic Solutions in Separable Form}
\label{sec:semi-ana}

\begin{figure}
  \centering
  \includegraphics[width=80mm,keepaspectratio]
  {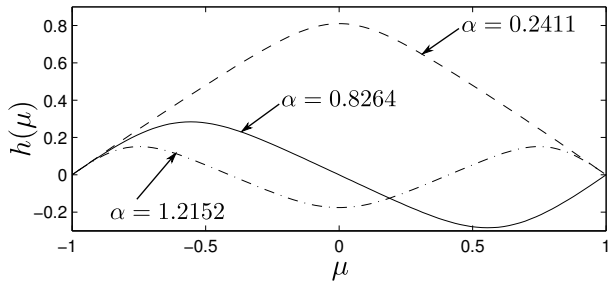}
  \caption{
  Examples of discrete numerical eigensolutions
   of second-order nonlinear ODE
   \eqref{eq:u-h-separated-ode} for $h(\mu)$.
  Dimensionless parameters are selected as $\beta_0 =
    \delta_0 = 0$, $\epsilon_0=10^{-2}$ and $\gamma_0=10^2$,
    while the normalization condition is chosen as
    $(\d h / \d\mu)|_{\mu=-1}=1$.
%{\bf How about a different value?}
  For a different normalization condition, the
  eigenfunction and eigenvalue are different
  yet with the same number of nodes.
  Distinct nonlinear eigensolutions are labelled by
  their corresponding eigenvalues of $\alpha$ here.
  Here, dashed curve for $\alpha=0.2411$, solid
    curve for $\alpha=0.8264$ and dash-dotted
    curve for $\alpha=1.2152$, respectively.}
  \label{fig:ODE_solution}
\end{figure}

Our first step is to numerically solve the separated
  nonlinear ODE \eqref{eq:u-h-separated-ode} of $h(\mu)$
  for semi-analytic solutions in the separable form.
The radial part as a factor in the separable form of expression
  \eqref{eq:u-separable-form} is simple once the power index
  $\alpha$ is given; we mainly focus on the latitudinal $\mu$
  part as a factor involving $h(\mu)$ for the angular $\theta$
  dependence.

\begin{figure}
  \centering
  \includegraphics[width=80mm,keepaspectratio]
  {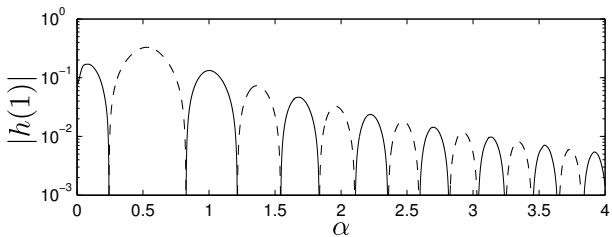}
  \caption{
  Examples of discrete numerical eigenvalues of
  second-order nonlinear ODE \eqref{eq:u-h-separated-ode}
  for $h(\mu)$.
  Pertinent dimensionless parameters are selected as $\beta_0
    =\delta_0 = 0$, $\epsilon_0=10^{-2}$ and $\gamma_0=10^2$,
    while the normalization condition is taken as
    $(\d h /\d\mu)|_{\mu=-1}=1$.
  We show here the plot of $|h(1)|$ (the absolute value
    of $h$ at $\mu=1$) versus different $\alpha$ values,
    where the curve is solid for $h(1) > 0$ and is dashed
    for $h(1) < 0$.
  Note that an $\alpha$ corresponds to an eigenvalue
    only when it makes $h|_{\mu=1} = 0$. }
  \label{fig:ODE_phase}
\end{figure}

By specifying the set of dimensionless parameters as
  $\beta_0 =\delta_0 = 0$, $\epsilon_0 = 10^{-2}$
%   {\bf Consistency of the last one?}
  and $\gamma_0=10^2$, and by applying the
  normalization condition of $(\d h /\d\mu)|_{\mu=-1}=1$,
  we present three examples of nonlinear eigensolution
  $h(\mu)$ in Fig. \ref{fig:ODE_solution}.
A different normalization corresponds to a different
  eigenfunction with different eigenvalue $\alpha$
  yet with the same node number in the latitudinal
  dimension $\mu=\cos\theta$.
Specifically, these nonlinear eigensolutions have different
  eigenvalues $\alpha$: the $n$th eigenvalue corresponds to
  an eigensolution with $(n+1)$ zeros or nodes
  in $\mu$ including the two poles where integer
  $n=1, 2, 3, \cdots$.
This leads to a set of the so-called
  ``nodal cones'' in the three-dimensional
  space for an axisymmetric MHD flow system.
For a clearer presentation, we show the eigensolutions
  of $h(\mu)$ corresponding only to the smallest three
  eigenvalues in succession in Fig. \ref{fig:ODE_solution}.
%  {\it
We further display the sequence of the smallest eigenvalues
%WLL  {\bf Meaning?}
   in Fig. \ref{fig:ODE_phase}, which
   illustrate how $h(1)\equiv h|_{\mu=1}$ varies for
   different $\alpha$ in the interval $0 <\alpha <4$.
It is natural to expect, especially in view of Fig.
  \ref{fig:ODE_phase}, that the number of eigenvalues may
  actually be very large or even (countable) infinite.
%  }.
%{\bf You may use the manner similar to that of figure 4
%  in Bruma \& Cuperman (1993) to show this trend empirically!}

Multiplied the latitudinal part $h(\mu)$ by the radial
  part $\eta^\alpha$, a separable semi-analytical
  solution $u(\eta,\ \mu)$ is then determined.
We show the $u(\eta,\ \mu)$ contour plot for one
  of them in subsection \ref{sec:not-unique} for comparisons,
  after directly solving PDE \eqref{eq:u-pde-dimensionless}
  with two distinctly different numerical codes of FDM and
  FEM yet with the same set of boundary conditions.

\subsection{Multiple MHD Solutions with the\\ \qquad
 Same Set of Boundary Conditions}
%\subsection{Are Those Semi-analytic Solutions Unique?
%              Alternatives?}
\label{sec:not-unique}

\begin{figure}
  \centering
  \includegraphics[width=80mm,keepaspectratio]
  {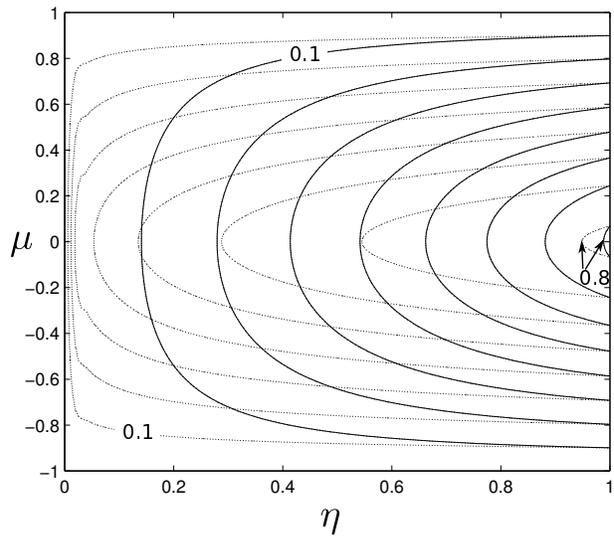}
  \caption{
    Comparisons between $u(\eta,\ \mu)$ contour plots of
      the semi-analytic solution with $\alpha=0.2411$,
      $\beta_0 = \delta_0 = 0$,
      $\epsilon_0=10^{-2}$ and $\gamma_0=10^2$ (i.e. the
      grey dotted contours), its corresponding FDM numerical
      results (black solid contours) and FEM numerical results
      (black dotted contours); the latter two sets of contours
      almost coincide and cannot be distinguished at this level.
  %     {\bf Need to check FDM and FEM. confirmed Feb 19, 2013.}
     Adjacent contours in each type are separated by value
       difference in $u$ of $0.1$, while the $u=0.1$ and $u=0.8$
       contour curves are labelled explicitly for each solution.
  %We note here that FDM and FEM contours are so nicely overlapped
  %   that they cannot be distinguished from each other.
       }
  \label{fig:Dipolar_contour}
\end{figure}

\begin{figure}
  \centering
  \includegraphics[width=80mm,keepaspectratio]
  {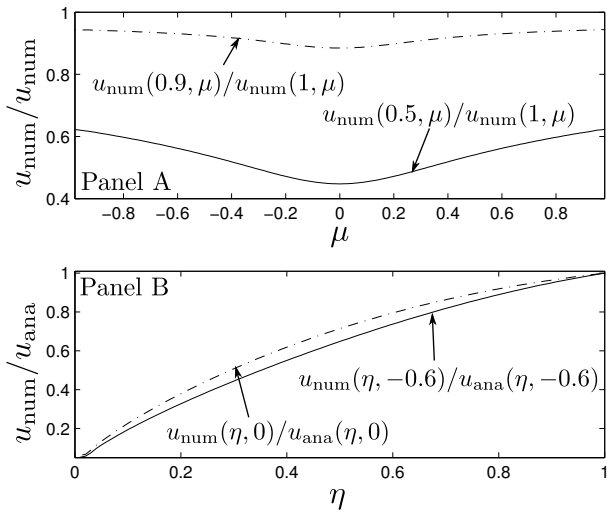}
  \caption{Here we compare the numerical solution
      $u_\mathrm{num}$ with its corresponding
      semi-analytic solution $u_\mathrm{ana}$ and
      with itself at different locations within
      the integration domain.
    Pertinent parameters adopted are $\alpha = 0.2411$,
      $\beta_0 = \delta_0 = 0$, $\epsilon_0 = 10^{-2}$
      and $\gamma_0 = 10^2$.
    Panel A shows the ratio of $u_\mathrm{num}
      (\eta_a,\ \mu) / u_\mathrm{num} (\eta_0,\ \mu)$,
      where $\mu\equiv\cos\theta$ is the variable argument
      and $\eta_a$ is a fixed $\eta$ value labelled in the
      figure ($\eta_a=0.9$ for the dash-dotted curve and
      $\eta_a=0.5$ for the solid curve).
    The curve variations instead of constants demonstrate
      that nonlinear solutions numerically solved by FDM
      and FEM are not separable (see the main text).
    Panel B shows the ratio of $u_\mathrm{num}
      (\eta,\ \mu_a) / u_\mathrm{ana} (\eta,\ \mu_a)$,
      where $\eta$ is the argument and $\mu_a$ is a
      fixed $\mu$ value ($\mu_a=0$ for the dash-dotted
      curve and $\mu_a=-0.6$ for the solid curve).
 %     {\bf Major points of such choices?}
      }
  \label{fig:Dipolar_comparison}
\end{figure}

In reference to our prior independent research work as described
  in Appendix \ref{sec:prev-works-multi}, it is of particular
  interest to examine whether the separable semi-analytic
  solutions to nonlinear PDE \eqref{eq:u-pde-dimensionless}
  are unique for the same prescribed boundary conditions
  (see subsection \ref{sec:sufficient-cond-unique} for
  a further elaboration).
In this subsection, we focus on the semi-analytic solution
  with $\alpha=0.2411$, $\beta_0 = \delta_0 = 0$,
  $\epsilon_0 = 10^{-2}$, $\gamma_0=10^2$ and
  $(\d h / \d\mu)|_{\mu=-1}=1$ (i.e. $n=1$ for the
  two-node eigensolution including the two poles in Fig.
  \ref{fig:ODE_solution}) as an example for illustration
  and elaboration.

Before running the codes of numerical PDE solvers,
  we prescribe the boundary condition as
  $u= h(\mu)|_{\alpha=0.2411}\eta_i^{0.2411}$ at
  $\eta=\eta_i=1$ and $u=0$ at $\eta\rightarrow 0^+$
  and at $\mu=\pm 1$ for the two poles.
We then substitute the same chosen forms of free
  functionals \eqref{eq:power-func-u} into nonlinear
  PDE \eqref{eq:u-pde-dimensionless} such that our
  numerical PDE solvers have been executed for the
  identical nonlinear PDE and boundary conditions as
  compared with separable semi-analytic solutions.
After running the two numerical FDM and FEM codes, our
  purely numerical solution results show many conspicuous
  differences from the separable semi-analytic ones.

In Fig. \ref{fig:Dipolar_contour}, we compare the
  separable semi-analytic solution (grey dotted contours),
  the FDM numerical solution (black solid contours) and
  the FEM numerical solution (black dotted contours).
While the direct computational results from the FDM and FEM
  codes show little (if any) differences, the difference
  between the two almost coincident numerical solutions
  and the separable semi-analytic one is conspicuous.
From the two sets of contours, we observe that at
  larger $\eta$ the numerical solutions drop faster for
  decreasing $\eta$ than the semi-analytic one does.
%WLL    {\bf Need a clarification!}
While the numerical solutions goes to zero almost at
  a constant rate, the semi-analytic one drops very
  slowly at first and dives very fast at last as
  $\eta$ goes from $\eta_i=1$ to $\eta_o=0$.

One might wonder whether the appearance of multiple
 nonlinear solutions (the key issue of non-uniqueness)
 may be somehow related to the boundary $\eta_o=0$ or
 $r\rightarrow +\infty$ where the $\mu$-dependence of
 the boundary condition becomes implicit and cannot
 manifest by degeneracy.
For this sake,
 we have actually explored further in the following manner.
For a range of small $\eta_o > 0$ close to $\eta_o = 0$,
 we specified appropriate semi-analytical solution as the
 boundary conditions at both $\eta_o$ and $\eta_i=1$.
%($\eta_i$  greater than $\eta_o$).
For both FDM and FEM numerical schemes, the numerical solutions
 consistently converge to corresponding solutions different from
 the initially specified semi-analytic solution;
these different convergent numerical solutions at different
 $\eta_o$ also differ among themselves.
For a sufficiently large $\eta_o$ (still smaller than $\eta_i=1$),
 we did the same by specifying appropriate semi-analytical solution
 as the boundary conditions at both $\eta_o$ and $\eta_i=1$.
For both FDM and FEM numerical schemes, the numerical solutions
 now converge to the initially specified semi-analytic solution.
This situation highlights the importance of outer boundary
 conditions (not just those at the stellar surface) in
 numerical simulations.
For example, one should be extremely careful in the contexts
 of numerical extrapolation of magnetic field configurations
 in solar/stellar coronae from the solar/stellar photospheric
 boundary conditions alone.
For astrophysical applications, we naturally emphasize more
 the limiting case of $\eta_o = 0$ for $r\rightarrow +\infty$.
The emergence of multiple possible nonlinear solutions (at
 least two by our exploration) with the same boundary
 conditions is therefore very interesting and raises
 important issues to be faced in several fronts.

For the purpose of removing suspicions that grid
 resolution might cause something unexpected numerically,
 we have also increased the grid resolutions for
 both the FDM and FEM numerical schemes for the
 same MHD model calculations.
The results of our numerical calculations and comparisons
 remain consistent and stable yet with longer running times.
With these extensive numerical explorations, the
 conclusions for the existence of nonlinear
 numerical solutions in addition to the corresponding
 semi-analytic solutions remain valid.

In Appendix B, we compared our numerical results
 with those of Bruma \& Cuperman (1993) for the
 problem of two-dimensional force-free magnetic
 field solutions.
As a matter of fact, when we first started our numerical
 explorations with both FDM and FEM schemes for the
 semi-analytic solutions of the force-free magnetic
 field problem by Low \& Lou (1990), we were not aware
 of the work by Bruma \& Cuperman (1993).
For such completely independent pursuits, their key
 numerical solutions and those of ours are remarkably
 consistent with each other (see Appendix B for more
 detailed information).

In our ideal MHD formalism, as long as the poloidal
 Alfv\'enic Mach number $M<1$ as defined in the
 paragraph before the integral transformation
 \eqref{eq:psi-u-transformation}, the elliptical
 nonlinear PDE \eqref{eq:u-pde-dimensionless} is well
 behaved everywhere and indeed we do not encounter any
 singularities or flow/field critical surfaces.
For MHD flows involving one or several MHD critical points,
 additional constraints need to be imposed for smooth solutions.
These additional requirements of crossing the MHD critical
 surfaces would be expected to fix nonlinear MHD solutions
 in unique manners.
Conceptually, these MHD critical points serve as ``boundary
 conditions" at smaller radii to warrant uniqueness of
 nonlinear numerical solutions.

More comparisons of different features are displayed in
  Fig. \ref {fig:Dipolar_comparison}, where we use the
  subscript ``num" for the numerical solution and ``ana"
  for the separable semi-analytic solution.
In Panel A, we show the ratio $u_\mathrm{num}
  (\eta_a,\ \mu)/u_\mathrm{num}(\eta_i,\ \mu)$, where
  $\eta_a$ takes a fixed $\eta$ value and $\mu$ is the
  variable argument of this ratio function (actually
  $u_\mathrm{num}(\eta_i,\ \mu)$ stands for the
  boundary condition at $\eta_i$).
Those two curves (at different $\eta_a$) in Panel A
  clearly shows that the numerical solution is not
  separable: should it be separable, this ratio
  would remain constant as $\mu$ varies.
Complementarily, Panel B presents $u_\mathrm{num}
  (\eta,\ \mu_a) / u_\mathrm{ana} (\eta,\ \mu_a)$, where
  $\mu_a$ is a fixed $\mu$ value and the variable
  argument of this ratio function is $\eta$.
The two curves with different $\mu$ show that the
  numerical solution falls faster than the separable
  semi-analytic one does.

%{\bf Stop here!}

\subsection{Flow Velocity and Magnetic Fields}
\label{sec:magnetic-lines-force}

\begin{figure*}
  \centering
  \includegraphics[width=180mm,keepaspectratio]
  {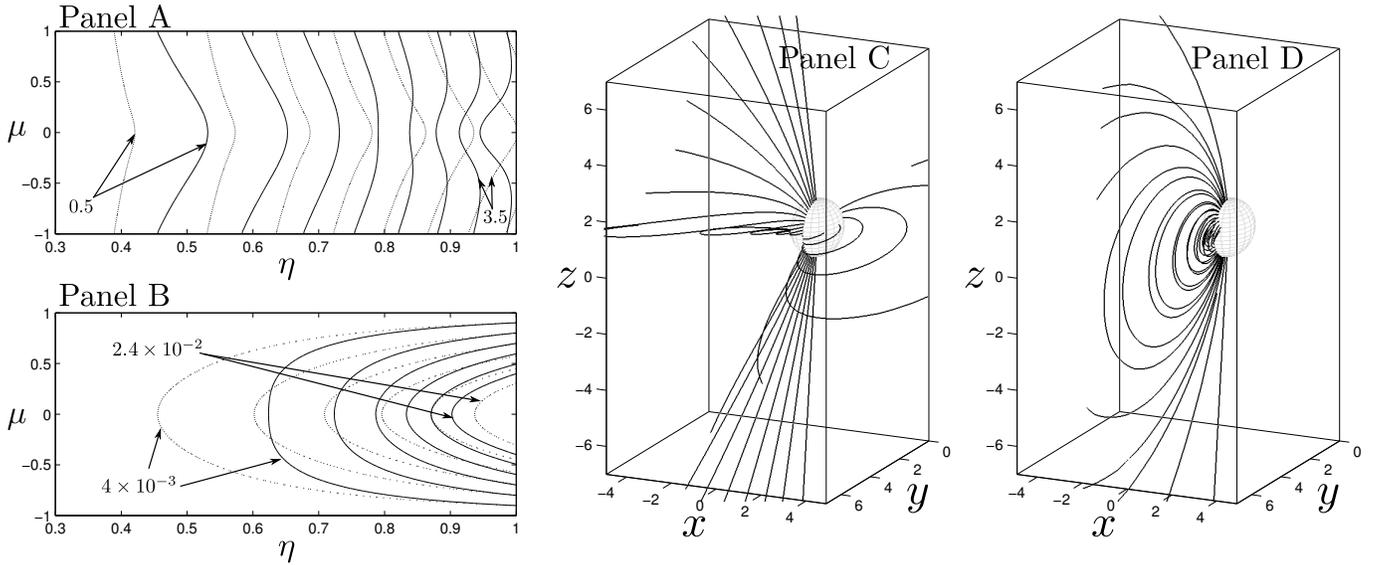}
  \caption{
    Presentation of magnetic and velocity fields generated
      from different types of steady MHD model solutions.
    Results are produced from the separable solution
     of nonlinear ODE \eqref{eq:u-h-separated-ode} with
     $\alpha=0.2411$,
     $\beta_0 = \delta_0 = 0$, $\epsilon_0=10^{-2}$,
     $\gamma_0=10^2$ and $(\d h/\d\mu)|_{\mu=1}=1$.
    Panel A displays the contour plots for the magnitudes
%      {\bf Need to check!}
     of dimensionless magnetic fields of semi-analytic
     solution (dotted contours) and numerical solution
     (solid contours).
    Adjacent contour curves are separated by an increment
      of $0.5$ in the magnitude of dimensionless magnetic
      field as function of $\eta$ and $\mu$.
    Panel B uses the same line types to present the
      respective magnitudes of dimensionless velocity fields,
      with adjacent contours being separated by $4\t10t{-2}$
      in the magnitude of dimensionless velocity as function
      of $\eta$ and $\mu$.
%    {\bf in what variable with an specific expression?}.
    Lines of magnetic (solid curves) and velocity (dotted
      curves) fields are shown in Panel C for the semi-analytic
      solution and Panel D for the numerical solutions.
    The initial points of 30 lines of force are evenly spaced
      on one specific meridian over the spherical surface of
      the central grey unit sphere.
    The meridian is located in the $x=0$, $y>0$ half plane.
    Note here that solid curves are so nicely overlapped by dotted
      curves that we cannot detect the differences by human eyes.
In this figure, $x$ stands only for the Cartesian coordinate, not
      the $x$ function as specified by free functional choice
      in equation \eqref{eq:power-func-u}. }
  \label{fig:Dipolar_b_v_field}
\end{figure*}

The most direct way to examine our 2.5D steady MHD results
  is to calculate the magnitudes of (dimensionless) magnetic
  and velocity fields and plot these field lines explicitly.
The method and related ansatz of constructing such field
  lines from dimensionless $u(\eta,\ \mu)$ function are
  detailed in Appendix \ref{sec:field-recover}.
Here, we show several results in Fig. \ref{fig:Dipolar_b_v_field}.

In Panels A and B of Fig. \ref{fig:Dipolar_b_v_field}, the
  magnitudes of magnetic and flow velocity fields versus
  $\eta$ and $\mu$ are shown in contour plots, respectively
  (solid contour curves for numerical solution and dotted
  contour curves for separable semi-analytic solution).
We note that the magnitudes of magnetic field and
  flow velocity field of the numerical solution fall faster
  than those of the separable semi-analytic solution do as
  $\eta\rightarrow 0^+$ corresponding to the radial range
  of very large $r$.

In Panels C and D of Fig. \ref{fig:Dipolar_b_v_field},
  lines of magnetic and flow velocity fields are displayed for
  the nonlinear separable semi-analytic solution (Panel C) and
  numerical solution (Panel D) respectively in dotted curves
  for flow velocity field and solid curves for magnetic field.
All field lines originate at the same meridian line on
  the central grey sphere (though they may terminate at
  somewhere else), which is sufficient to sketch the
  global configuration as the axisymmetric system is
  invariant under a rotation about the $z$ axis.
Although we plot the lines of magnetic and flow velocity fields
  in different line styles, the two types of lines overlap
  very well, which reveals that flow of fluid particles are
  closely parallel or anti-parallel to the magnetic lines
  of force.
%  {\bf This coincidence expected theoretically?}
This confined parallel or anti-parallel motions are
  expected if we note $\delta_0=0$ in this model;
% this $\delta_0$ indicates that the electric
%   field vanishes, implying that the field is force-free.
%  {\it
    this $\delta_0$ parameter indicates that the electric
    field vanishes, implying $\bvec\times\vvec = 0$ from
    $\evec + \vvec\times\bvec/c=0$ for ideal MHD equation
    \eqref{eq:mhd-equ}.
%   }
%  {\bf Why is this claim?}{\it I made an error then; should
%    not be force-free.}
From Fig. \ref{fig:Dipolar_b_v_field}, one can readily tell the
  difference in the field configurations between those two types
  of solution in a very intuitive manner: toroidal components
  generated by numerical solution is much weaker than that
  generated by the separable semi-analytic solution.

%  {\bf Stop here!}

\subsection{Further Numerical Explorations\\
  \qquad\ by Our FDM and FEM Solvers}
\label{sec:more-exploration}

\subsubsection{Presentation of Numerical Results}
\label{sec:presentation-results}

With the numerical ``pipeline" established above for
  either FDM or FEM codes, we can readily construct
  MHD configurations of magnetic and flow velocity
  fields to examine their various global features.
This subsection offers an example, where we solve
  nonlinear ODE \eqref{eq:u-h-separated-ode} and
  nonlinear PDE \eqref{eq:u-pde-dimensionless}
  with the following conditions and parameters:
%  {\bf Sufficiently complete?}
%   {\it (these parameters are complete)}
  $(\d h /\d\mu)|_{\mu=-1}=2$, $\alpha=0.4311$,
  $\beta_0=10^{-8}$ (almost zero), $\gamma_0=10^2$
  and $\delta_0 = \epsilon_0 = 10^{-3}$.
Solution of nonlinear ODE \eqref{eq:u-h-separated-ode} is
  presented in Panel A of Fig. \ref{fig:Quadrupole_contour},
  while the comparison between the semi-analytic solution
  and the numerical solutions (still with both FDM and FEM
  calculations) is displayed in Panel B.
The situation in Fig. \ref{fig:Quadrupole_contour}
  is qualitatively similar to that in Fig.
  \ref{fig:Dipolar_contour}: the numerically obtained PDE
  solution falls faster than the semi-analytic one does
  as $\eta$ goes to zero in the regime of very large $r$.

For the convenience of reference, we summarize the chosen
  parameters and notations for our dipole and quadruple
  steady MHD models in Table \ref{table:parameters-model}.

\begin{table}
  \centering
  \caption{
    We summarize 2.5D steady MHD model parameters below.
    The notations D$_\mathrm{num}$ and D$_\mathrm{ana}$ refer to
     respectively the numerical and semi-analytic models with
     dipolar configuration in Fig. \ref{fig:Dipolar_contour}.
    Notations Q$_\mathrm{num}$ and Q$_\mathrm{ana}$, on the
    other hand, denote the numerical and semi-analytic MHD
    models with quadrupolar configuration (see Fig.
    \ref{fig:Quadrupole_contour} for more details). }
  \begin{tabular}{ccc}
    \hline
    Parameters & D$_\mathrm{num}$ and D$_\mathrm{ana}$
       & Q$_\mathrm{num}$ and Q$_\mathrm{ana}$ \\
    \hline
    $\alpha$ & 0.2411 & 0.4311 \\
    $\beta_0$ & 0 & $10^{-8}$ \\
    $\gamma_0$ & $10^2$ & $10^2$ \\
    $\delta_0$ & 0 & $10^{-3}$ \\
    $\epsilon_0$ & $10^{-2}$ & $10^{-3}$ \\
    $(\d h/\d\mu)_{\mu=-1}$ & 1 & 2 \\
    \hline
  \end{tabular}
  \label{table:parameters-model}
\end{table}

\subsubsection{Magnetic and Flow Velocity Fields}
\label{sec:magn-veloc-field}

Fig. \ref{fig:Quadrupole_b_v_field} is generated by the
  scheme identical to Fig. \ref{fig:Dipolar_b_v_field}.
From Panels A and B of Fig.
 \ref{fig:Quadrupole_b_v_field}, we acquire similar
  information about the magnetic and flow velocity fields
  as in subsection \ref{sec:magnetic-lines-force}, although
  the distributions seem to be full of features in their
  characteristics.
In addition, we observe that the magnitude of flow
  velocity is fairly small near the equatorial plane
%  {\bf Explanations?}
  since the values of $u$,
  $\partial u/ \partial \eta$ and other quantities
  proportional to $u$ and  $\partial u/ \partial \eta$ vanish
  at the equatorial plane, implying that the $\theta$ and
  $\phi$ components are small near the equator $\mu=0$
  (see Appendix \ref{sec:field-recover} for details).

\begin{figure}
  \centering
  \includegraphics[width=80mm,keepaspectratio]
  {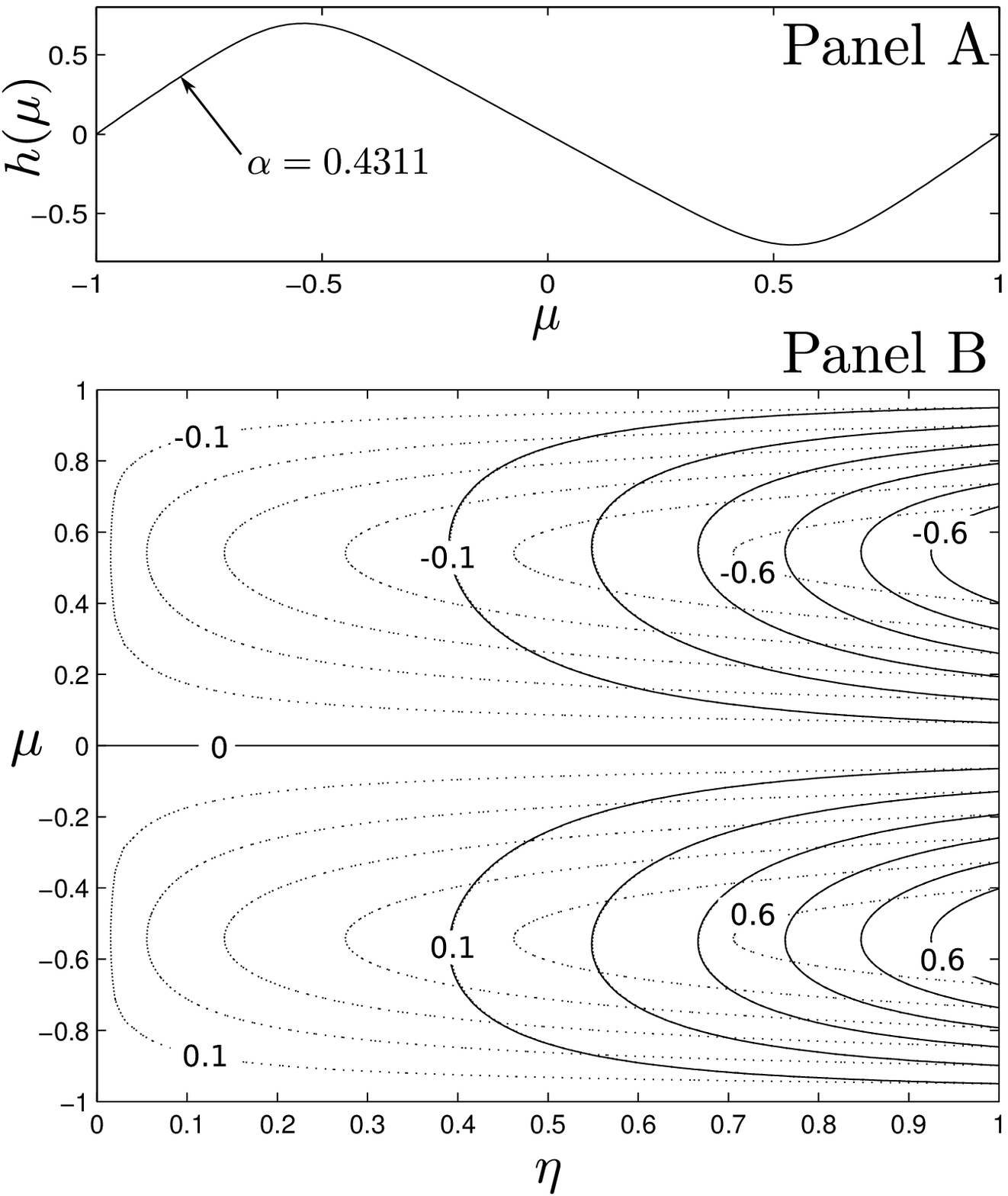}
  \caption{
    Solutions of $u(\eta,\ \mu)$ from nonlinear PDE
      \eqref{eq:u-pde-dimensionless} with a
      quadrupole configuration.
    Panel A gives the separable eigensolution of nonlinear ODE
      \eqref{eq:u-h-separated-ode} with pertinent parameters
      for Q$_\mathrm{ana}$ and Q$_\mathrm{num}$ as given in
      Table \ref{table:parameters-model}.
    This eigensolution gives the separable semi-analytic
      solution by equation \eqref{eq:u-separable-form}.
    Panel B compares the contour plots of the
      semi-analytic solution generated from Panel A (grey
      dotted contours), the corresponding FDM numerical
      results (black solid contours) and the FEM
      numerical results (black dotted contours).
    Here again, FDM and FEM contours almost coincide
      with each other and cannot be distinguished
      from each other.
    Adjacent contour curves of each type are separated by
      $0.1$, while the $u=0$, $u=\pm 0.1$ and $u=\pm 0.6$
      contours are marked explicitly for each solution. }
  \label{fig:Quadrupole_contour}
\end{figure}

\begin{figure*}
  \centering
  \includegraphics[width=180mm,keepaspectratio]
  {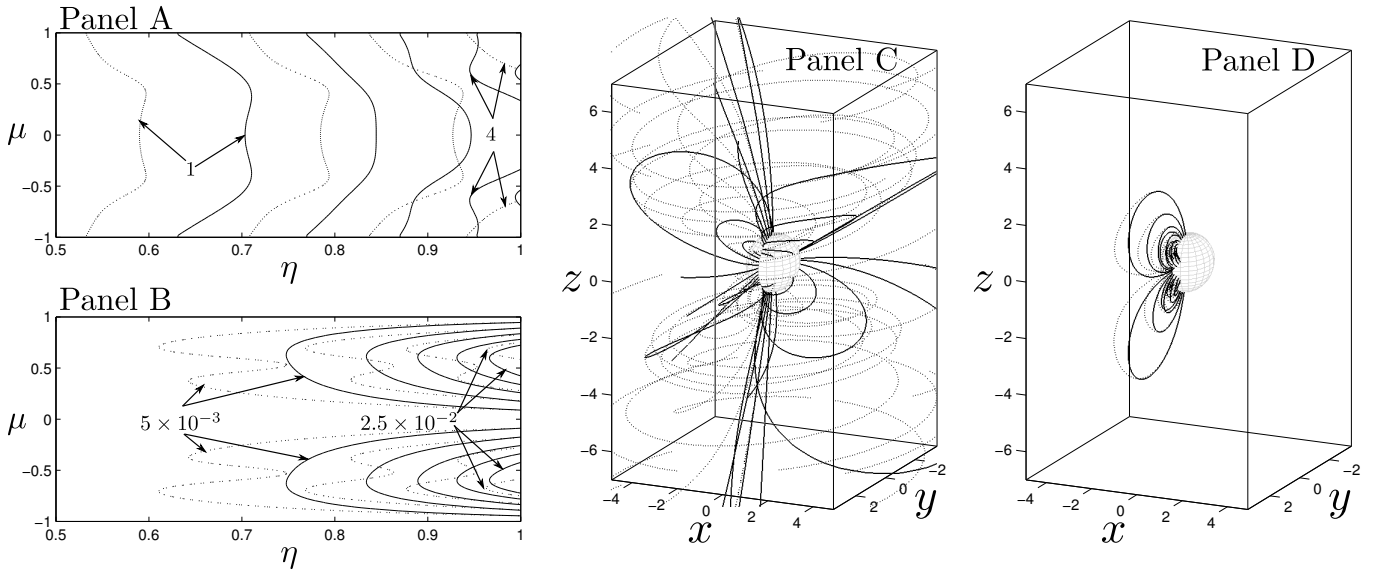}
  \caption{
    The quadrupole magnetic and flow velocity fields are shown here.
    Results are generated from the separable semi-analytic solution
     and direct numerical solutions shown in
     Fig. \ref{fig:Quadrupole_contour}.
    Panel A is the contour configuration showing the magnitude
      of dimensionless magnetic field versus $\eta$ and $\mu$
      of semi-analytic solution in dotted contours and numerical
      solution in solid contours.
    Adjacent contour curves are separated by an increment
      of $1.0$ in the magnitude of dimensionless magnetic
      field as function of
  %  {\bf Sure of this?} in terms
      of $\eta=r_0/r$ and $\mu=\cos\theta$.
%    {\bf in what variable?}
    Panel B uses the same respective line types to display the
      magnitude of dimensionless flow velocity field, with
      adjacent contours being separated by $5\t10t{-3}$ in
      the magnitude of dimensionless velocity as function
%   {\bf Sure of this?} in terms
      of $\eta$ and $\mu$.
%      {\bf in what variable?}.
      Lines of magnetic field (solid curves) and flow
      velocity (dotted curves) field are shown in
      Panel C for the semi-analytic solution and
      Panel D for the full numerical solution.
    Those field lines are generated with the same scheme
      as done in Fig. \ref{fig:Dipolar_b_v_field}. }
  \label{fig:Quadrupole_b_v_field}
\end{figure*}

The nonlinear ODE solution has three zero points
  or nodes (this would be impossible
  if we do not generalize the ansatz of
  \citealt{2001GApFD..94..249T} by taking
  the absolute value of $u$ in the selection
  of free functionals);
%   {\bf Please explain still!}
   the magnetic field shows a different topology -- it
   takes a quadrupole configuration which is shaped by the
   angular (i.e. $\theta$ or $\mu$) dependence of $u$
   function and the dependence of magnetic field on
   $u$ (see Fig. \ref{fig:Quadrupole_contour} and
   Appendix \ref{sec:field-recover}).
%  {\bf Please explain.}
However, according to Panels C and D of Fig.
  \ref{fig:Quadrupole_b_v_field}, we can see more
  features of this 2.5D steady MHD solution.
%   {\bf Wording?}
Both panels show that the flow velocity field lines have
  apparently different configurations in reference
  to the magnetic lines of force,
  % i.e. the fluid
  % particles have much more ``freedom''.
%  {\it
  i.e. the impact of macroscopic electric field
  is not negligible under those conditions.
%  }
%  {\bf Please clarify this perspective
%   in terms of rotation! Plasma remains attached
%   to the magnetic field lines.}
% {\it
We attribute this phenomenon to the interaction between the
   plasma mass flow, the magnetic field, and the rotation
   of the central stellar object.
%  }
In Panel D, corresponding to the numerical solution, flow
  velocity field lines, more or less, look similar to the
  magnetic field lines, showing the fact that fluid particle
  motions are almost parallel or anti-parallel to magnetic
  field lines near the central stellar object in general.
Nevertheless in Panel C, the scenario appears quite
  different.
While the magnetic field lines show a distorted quadrupole
  field configuration, the flow velocity field manifests
  an apparent helical shape.
Fluid particles travel along spiral lanes, escaping from
  the vicinity of the central stellar object to infinity.
Magnetic field is still important to fluid particles,
  but electric field are comparable (see subsection
  \ref{sec:dimension-recovering} for details; here,
  $\delta_0$ is set to be small yet sufficiently
  significant).
This may happen if we consider the equation
  $\evec = -\vvec\times\bvec/c$
%{\bf Minus sign missing? Sign correction confirmed with Lile.}
  when the flow velocity $\vvec$ and the magnetic field
  $\bvec$ do not align with each other.
The scenario here tells a different story in contrast to
  the case of subsection \ref{sec:magnetic-lines-force}
  where the magnetic and velocity fields have field lines
  with the same configurations.
%  {\bf Please explain explicitly!}
%  {\bf  Magnetic and velocity fields? Meaning of
%``almost"?} {\it ``almost'' deleted }

\section{Analyses and Applications}
\label{sec:applicatoins}

\subsection{Multiple Nonlinear Solutions}
\label{sec:discuss-multiplicity-energy}

\subsubsection{A sufficient condition for a
  quasi-linear\\
\qquad\quad  PDE to possess unique solution}
  \label{sec:sufficient-cond-unique}

For a nonlinear elliptical PDE, the uniqueness
  of its solutions is not guaranteed for the
  same boundary conditions in general.
In this subsection, we specifically focus on a
  functional property associated with nonlinear
  PDE \eqref{eq:u-pde-dimensionless}.

For the convenience of analysis, we first recast nonlinear PDE
  \eqref{eq:u-pde-dimensionless} in the following form of
\begin{equation}
  \label{eq:u-pde-standardform}
  0 = \dfrac{1}{(1-\mu^2)}\dfrac{\partial}{\partial
    \eta}\left( \eta^2 \dfrac{\partial u}{\partial \eta}
  \right) + \dfrac{\partial^2u}{\partial \mu^2} +
  \dfrac{\mathcal{F}(u,\ \eta,\ \mu)} {\eta^2(1-\mu^2)}\ ,
\end{equation}
where the functional $\mathcal{F}(u,\ \eta,\ \mu)$
 is explcitly defined by
\begin{equation}
  \label{eq:f-func-standardform}
  \begin{split}
    \mathcal{F}(u,\ \eta,\ \mu)
    & \equiv \dfrac{\alpha^2 \gamma_0}{2}
    \dfrac{\d}{\d u}(x^2)+\dfrac{\alpha^2\beta_0}{2}
    \left( \dfrac{1-\mu^2}{\eta^2} \right)
    \dfrac{\d p_s}{\d u}
    \\
    & \qquad + \alpha^2 \delta_0 \left( \dfrac{1-\mu^2}
      {\eta^2}\right)^2 \dfrac{\d}{\d u}\left[ \varrho
      \left( \dfrac{\d \varphi}{\d u}\right)^2 \right]
    \\
    & \qquad + \dfrac{\alpha^2 \epsilon_0}{2}
      \left(\dfrac{1-\mu^2}{\eta}\right)
      \dfrac{\d \varrho}{\d u} \
  \end{split}
\end{equation}
which further contains several free functionals.
 The problem of solving nonlinear PDE
  \eqref{eq:u-pde-standardform} may be then cast
  into a functional analysis problem, where we
  seek the minimum of the following functional
  with prescribed boundary conditions
%  \citep[see also][for how and why to obtain a
%  functional in such a form]{2011PDE..book.Taylor}:
\begin{equation}
  \label{eq:u-pde-functional}
  \begin{split}
    J(u) = & \dfrac{1}{2} \iint \d \mu \d \eta \bigg[
    \dfrac{\eta^2}{(1-\mu^2)} \left( \dfrac{\partial u}
      {\partial \eta}\right)^2 +\left(\dfrac{\partial u}
      {\partial \mu} \right)^2
    \\
    & \qquad\qquad\qquad\qquad - 2\int_0^u
    \dfrac{\mathcal{F}(u'\! ,\ \eta,\ \mu)}
    {\eta^2(1-\mu^2)} \d u'
    \bigg]\ .
  \end{split}
\end{equation}
% {\bf What is the basic reason for such a choice of integral?}
The interested reader is referred to the book of
  \cite{2011PDE..book.Taylor} for how and why to obtain
  a functional of such a form in more specific details.
Mathematically, the minimum of functional
  \eqref{eq:u-pde-functional} is unique when the functional is
  strictly convex, or equivalently, the following inequality
  needs to be satisfied \citep[see][]{2011PDE..book.Taylor}, namely
\begin{equation}
  \label{eq:unique-condition}
  \begin{split}
    0 \leq & - \dfrac{\partial^2}{\partial u^2} \left[ 2
      \int_0^u \dfrac{\mathcal{F}(u'\! ,\ \eta,\ \mu)}
      {\eta^2(1-\mu^2)} \d u '\right]
    \\  \qquad\qquad\qquad
    = & \dfrac{-2}{\eta^2(1-\mu^2)}
    \dfrac{\partial}{\partial u} \mathcal{F}(u,\ \eta,\ \mu) \ .
  \end{split}
\end{equation}
When all the free functionals of
  $u$ as specified by equation \eqref{eq:power-func-u} are
  prescribed, it is straightforward to verify that inequality
  (\ref{eq:unique-condition}) is not satisfied.
In fact, the chosen functionals by expression
  \eqref{eq:power-func-u} give a right-hand side (RHS)
  of equation \eqref{eq:unique-condition} that not
  only is negative definite but also does not have
  a lower bound as well.
% {\bf Please show this more explicitly.}
It is straightforward to demonstrate this by noting
  that the following equation is not bounded
  [for $\sigma > 2$, which holds true for all terms
  in equation \eqref{eq:f-func-standardform}]:
  \begin{equation}
    \label{eq:positive-indefinite}
    \dfrac{\d^2 |u|^\sigma}{\d u^2}
    = \sigma(\sigma-1) |u|^{\sigma-2}\ ,
  \end{equation}
  which describes a typical case for each term
  in equation \eqref{eq:f-func-standardform}.
We also note that all coefficients in front of the
  chosen functionals of $|u|$ in expression
  \eqref{eq:f-func-standardform} are positive definite.
Therefore the possibility cannot be excluded a priori
  that nonlinear PDE \eqref{eq:u-pde-dimensionless}
  together with the violation of inequality
  \eqref{eq:unique-condition} may possess multiple
  nonlinear solutions with the same boundary conditions.
We reiterate that the violation of inequality
  \eqref{eq:unique-condition} is not a proof that
  multiple solutions do exist necessarily; it only
  indicates that such possibility might occur.
We mainly rely on the following empirical fact:
  to the extent of the reliability of extensive numerical
  experiments by the FDM and FEM solvers, our numerical
  cross-checking calculations do reveal at least an
  alternative non-separable nonlinear solution exists
  corresponding to each separable nonlinear
  semi-analytic solutions.
%   which is verified by our revelation of multiple
%   solutions by the numerical integration.
In particular, one should pay considerable attention
  to the important role of outer boundary conditions
  at large radii for extrapolating magnetic field
  configurations in solar/stellar coronae and
  magnetospheres.

\subsubsection{Energies Associated with the Nonlinear Solutions}
\label{sec:mag-kinetic-energy}

According to the numerical calculations, even if we start the FDM
  relaxation from initially specified separable semi-analytic solutions,
  the numerical iterations usually converge to corresponding nonlinear
  solutions, as shown in both Figs. \ref{fig:Dipolar_contour}
  and \ref{fig:Quadrupole_contour}.
Based on our extensive numerical explorations, we conjecture that
  the direct numerical solutions might have lower energies than
  the corresponding semi-analytic separable solutions do such
  that the eventual convergence becomes more stable.
%{\bf Necessarily so?}{\it Not definitely sure.}

The dimensionless magnetic and kinetic energies are defined
  respectively by [see equation \eqref{eq:fiducial-velocity}
  in Appendix \ref{sec:field-recover} for the explicit
  definition of $v_0$]
\begin{equation}
  \label{eq:energy-definition}
  E_B = \dfrac{1}{B_0^2}\int_V \dfrac{\bvec^2}{8\pi} \d V\ ,\qquad E_K
  = \dfrac{1}{\rho_0v_0^2}\int_V \dfrac{\rho \vvec^2}{2} \d V\ ,
\end{equation}
where the dimensionless volume $V$ occupies the
  space outside the central stellar object and
  $\d V=2\pi \sin{\theta} d\theta r^2 dr /r_0^3$
  is the dimensionless differential volume element.
%  {\bf Please clarify specifically!}

We compute the field energy carried by the
  MHD flow configuration numerically.
For the dipolar MHD flow model D$_\mathrm{ana}$
  and D$_\mathrm{num}$, we have
\begin{equation}
  \label{eq:dipole-dimensionless-eng}
  \begin{split}
    & E_{B,\mathrm{ana}} = 4.92\ ,\qquad\ \qquad
      E_{K,\mathrm{ana}} = 6.28\t10t{-6}\ ;
    \\
    & E_{B,\mathrm{num}} = 3.04\ ,\qquad\qquad
      E_{K,\mathrm{num}} = 3.43\t10t{-6}\ .
  \end{split}
\end{equation}
%WLL {\bf How about the gravitational energy?}
For the quadrupolar MHD flow model Q$_\mathrm{ana}$
  and Q$_\mathrm{num}$, we have accordingly
\begin{equation}
  \label{eq:dipole-dimensionless-eng}
  \begin{split}
    & E_{B,\mathrm{ana}} = 4.11\ ,\qquad\qquad
      E_{K,\mathrm{ana}} = 2.69\t10t{-6}\ ;
    \\
    & E_{B,\mathrm{num}} = 2.09\ ,\qquad\qquad
      E_{K,\mathrm{num}} = 2.04\t10t{-6}\ .
  \end{split}
\end{equation}
%WLL   {\bf How about the gravitational energy?}
From the above results, we clearly observe that the
  numerical models presented in this paper have
  significantly, sometimes nearly a half, less
  magnetic energy than the semi-analytic ones do.
This might explain why our relaxation results so
  far always converge to the alternative nonlinear
  solutions with lower energies.
%{\bf Sensible reasoning? talked to Lile Feb 21, 2013}
However, it is not yet clear whether there exists a local
  minimum of energy around the semi-analytic solutions,
  which would then render them physically plausible as well.

In order to obtain dimensional results for both the
  magnetic and kinetic energies, we simply apply
  the following conversion of physical parameters:
\begin{equation}
  \label{eq:dimensional-energy}
  \mathcal{E}_B = E_B B_0^2r_0^3\ ,
  \qquad\qquad \mathcal{E}_K = E_K
  \dfrac{2G\mathcal{M}\rho_0r_0^2}{\epsilon_0}\ .
\end{equation}
Based on the quadrupolar model (Q$_\mathrm{ana}$
  and Q$_\mathrm{num}$)
  and the dimensional transformation in subsection
  \ref{sec:dimension-recovering}, we might expect a physical
  transition (say, in the presence of inevitable resistivity
  and/or viscosity -- e.g., Low 2013)
  from the configuration of semi-analytic model to that of the
  corresponding numerical model since they have the identical
  distribution of the normal magnetic field component on the
  spherical surface of the central stellar object.
Should that happen, the magnetic and kinetic energy releases
  would be given by
%  {\it [we adopt the physics parameters for numerical
%  quadrupolar Q$_\mathrm{num}$ results as in subsection
%  \ref{sec:dimension-recovering}, and especially equation
%  \eqref{eq:dimensionless-wd}] }
\begin{equation}
  \label{eq:dimensional-delta-energy}
  \quad\Delta \mathcal{E}_B \simeq 2\t10t{37}\erg\ ,\ \qquad
  \Delta \mathcal{E}_K \simeq 2\t10t{30}\erg\ ,
\end{equation}
respectively, where we adopt the physical parameters
   for numerical quadrupolar Q$_\mathrm{num}$ results
   in subsection \ref{sec:dimension-recovering}, and
   especially in reference to equation
   \eqref{eq:dimensionless-wd}.
% {\bf What are the parameters adopted for these numbers?}
It is apparent that the main portion of energy in the stellar
  magnetosphere is contained within the magnetic field, even
  though the physical conditions are not ``extreme'' at all
  (e.g. the magnetosphere of a magnetic white dwarf as in
  subsection \ref{sec:dimension-recovering}).

\subsection{Stellar Magnetospheres}
\label{sec:magnetosphere}

\subsubsection{Magnetic Field Constraints on Plasma Flows}
\label{sec:magnetic-constrain}

In cases that the magnetic field is overwhelmingly predominant
  in a stationary MHD equilibrium, partialy ionized plasma are
  almost strictly confined along the magnetic field lines.
However, when the misalignment between the directions of
  bulk flow velocity $\vvec$ and magnetic field $\bvec$ become
  significantly important, the role of the electric field
  $\evec$ (perpendicular to both $\vvec$ and $\bvec$) figures
  prominently and more diverse MHD features may emerge.
A diagnostic about the extent of plasma confinement
  is required to distinguish these two types of
  situations qualitatively.
%WLL   {\bf Be careful!}
For this purpose, we introduce the Cauchy-Schwartz correlation
  function \citep[e.g.][]{2006A&A...446..691A} below, namely
\begin{equation}
  \label{eq:c-s-correlation}
  \cscor = \dfrac{1}{V} \int_V
  \dfrac{(\bvec/B_0)\cdot(\vvec/v_0)}
  {|\bvec/B_0| |\vvec/v_0|} \d V\ ,
\end{equation}
where $V$ stands for the dimensionless spatial volume
  for evaluating the Cauchy-Schwartz correlation (i.e.
  the entire spatial domain outside the central spherical
  stellar object in the present contexts) and $\d V$ is
  the dimensionless differential volume element;
  subscripts $_{cs}$ is
  related to the initials of Cauchy-Schwartz.
%Although reducing the dimensions is not necessary,
We introduce fiducial values in order to give
  clearer expressions which are dimensionless.
It is obvious that the range of $\cscor$ values is
  $|\cscor|\leq 1$ and only when the two vector fields
  $\vvec$ and $\bvec$ are parallel or anti-parallel to
  each other everywhere, we would achieve the extremum
  values of $\cscor=\pm 1$.
For example, $\cscor$ should be fairly close to
  $1$ if the misalignment between the two vector fields
  $\vvec$ and $\bvec$ remains insignificant everywhere.

More specifically, we compute values of $\cscor$ for all four
  MHD solution models listed in Table \ref{table:parameters-model}.
Results are shown below (n.b. the relative numerical precision
  of our computations is $\sim 10^{-16}$, i.e. any results showing
  1 should be some value larger than $1-10^{-16}$):
%WLL    {\bf Not precise!}
\begin{equation}
  \label{eq:cs-cor-for-models}
  \begin{split}
    & \cscor(\mathrm{D}_\mathrm{ana}) > 1 - 10^{-16}\ ,
    \\
    & \cscor(\mathrm{D}_\mathrm{num}) > 1 - 10^{-16}\ ,
    \\
    & \cscor(\mathrm{Q}_\mathrm{ana}) = -0.255883\ ,
    \\
    & \cscor(\mathrm{Q}_\mathrm{num}) = 0.966990\ .
  \end{split}
\end{equation}
%  {\bf The inconsistent statements around need to be checked!}
It is clear that $|\cscor(\mathrm{Q}_\mathrm{ana})|$ is
  considerably smaller than the other three, indicating that
  the fluid particles tend to move nearly perpendicular to the
  magnetic field lines and to maintain an electric field.
Panel C of Fig. \ref{fig:Quadrupole_b_v_field} illustrates
  this situation transparently, especially around the two polar
  zones, where flow velocity lines go up helically around the polar
  axis while the magnetic field lines point nearly straight outwards.
Here, the rotation of the central stellar object plays an
  important role (see also subsection \ref{sec:flow}).
%   {\bf Implying a rotation? Mentioned to Lile.}
In comparison, $|\cscor(\mathrm{Q}_\mathrm{num})|$ is
  considerably larger than $|\cscor (\mathrm{Q}_\mathrm{ana})|$,
  corresponding to the fact, as can be seen in Panel D of Fig.
  \ref{fig:Quadrupole_b_v_field}, that the flow velocity field
  lines are distorted yet remain similar to the configuration
  of magnetic field lines.
%  {\bf Please emphasize the role of rotation more fully.}
%{\it
Nonlinearity of this MHD problem is hereby illustrated
  by examples.
While the rotational features of the central stellar objects,
  to which we attribute the initiation of toroidal fields, are
  the same for
  $|\cscor(\mathrm{Q}_\mathrm{num})|$ and $|\cscor
  (\mathrm{Q}_\mathrm{ana})|$ (see subsection \ref{sec:flow}), the
  interaction between the plasma mass flux and the magnetic field
  can manifest much different patterns of field lines.
%}

\subsubsection{Dimensional Recovery and Physical Conditions}
\label{sec:dimension-recovering}

Given considerable idealizations, theoretical
  MHD models in this formalism should be checked for
  the plausible utility by their abilities to describe
  magnetospheres of stellar objects in astrophysics.
In such applications, all dimensionless parameters defined
  in equation \eqref{eq:dimensionless-notes} for a model
  formulation need to be specified {\it a priori} as they
  directly determine the properties of the chosen MHD
  model framework.
Here, we offer explanations for our choice
  of pertinent physical parameters.

For a field configuration with considerable toroidal
  component, $\gamma_0$ parameter, which reflects
  the significance of a toroidal field component,
  should be somehow
%  {\bf toroidal magnetic field and rotation?}
  sufficiently large, but with a certain
  level of arbitrariness.
This point is more clearly demonstrated by noting
  equation \eqref{eq:scalar-relation} that $X(\psi)$
  combines the contributions of $I$, $F'$ and $\Phi'$,
  where $I$ indicates the toroidal component of the
  magnetic field.
An adopted value of $\gamma_0\sim 10^2$
%{\bf , corresponding to what physically for this choice?},
  corresponds to a configuration whose toroidal
  magnetic field component is extremely strong.

For the remaining parameters, we focus on the possibility
  of using our theoretical model to describe magnetospheric
  configurations in the context of magnetic white dwarfs
  (MWDs) as an example of astrophysical applications
% (e.g. Lou 1995).
  \citep[e.g.][]{1995MNRAS.276..769L,2009MNRAS.396..878H,
    1992ARA&A..30..143C,2005MNRAS.356..615F,2006MNRAS.367.1323F,
    2008MNRAS.389L..66F}.
%  {\bf Also references of Hu \& Lou 2009 on magnetars,
%  Chanmugam G. 1992, ARA\&A, Ferrario L. \& Wickramasinghe D. T.
%  2005, 2006, 2008 and so forth on magnetic white dwarfs. }
Those small earth-size compact degenerate stars usually have
  mass and radius in the orders of $\mathcal{M}\simeq 1M_\odot$
  (the Chandrasekhar upper mass limit of $1.4M_\odot$) and
  $r_0\simeq 10^9\mathrm{\ cm}$.
The magnetic field strengths of MWDs usually fall in
  the approximate range of $10^4\sim 10^9 \gauss$
  \citep[e.g.][]{1995ApJ...448..305S}.
Comparing to typical magnetic field strengths of neutron
  stars (e.g., magnetars, AXPs, radio pulsars, isolated
  neutron stars, millisecond pulsars etc.), this range
  for MWDs is relatively weak.
As an example, we take $B_0 = 10^5 \gauss$ as ``typical''
  magnitude of magnetic field strength in our model for a MWD
%  {\bf
  The rotation periods of white dwarfs can be days or
  even years.
% ; please check the references.}{\it Checked.}
%We note also here that, even f
  For white dwarfs, there are still possibilities
  that the magnetic field could be much stronger.
For example, some cataclysmic variables (CVs) may have
  their magnetic fields as strong as $\sim 10^8\gauss$
  \citep[e.g.][]{1977ApJ...216L..45K}.
%  {\bf This value could be chosen larger!}

As for a possible MWD wind flow, other relevant physical
  quantities are taken from \citet{1998ApJ...503L.151L},
  where particle number density is
  $n\sim 10^{13}\mathrm{\ cm}^{-3}$ and
  temperature is about $T\sim 6\t10t{6}\mathrm{\ K}$
  (these values are calculated at the zero optical
  depth radius for the chromosphere of a MWD).
The fiducial value of mass density is thus given by
  $\rho_0\simeq n m_\mathrm{p}\sim 10^{-11}\mathrm{\ g\
  cm}^{-3}$ ($m_\mathrm{p} = 1.67\t10t{-24}\mathrm{\ g}$
  is the proton mass).
The typical static magnetospheric plasma pressure $P_{s0}$
  may be possibly estimated by the ideal gas law as
  $P_{s0}\simeq
  n k_\mathrm{B}T \sim 10\mathrm{\ dyn\ cm}^{-2}$
  where $k_\mathrm{B}$ is the Boltzmann constant.
With all of these parameter estimates into
  definition \eqref{eq:dimensionless-notes},
  we would approximately have
\begin{equation}
  \label{eq:dimensionless-wd}
  \qquad\qquad \epsilon_0 \sim 10^{-3}\
  \qquad\hbox{ and }\qquad \beta_0 \sim 10^{-8}\ .
\end{equation}
These choices of two dimensionless parameters indicate
  that the gas pressure is dynamically insignificant,
  and so is the gravitational potential energy density,
  as compared to the magnetic pressure.
% {\bf Physical meanings of these two parameters.}
% {\bf You may search and read more papers on magnetic white
%  dwarfs, as well as the so-called cataclysmic variables (CVs)
%  -- the white dwarf component can be strongly magnetized. }

Coming to parameter $\delta_0$, things could become
  somewhat subtle.
The first braces in the expression of $\delta_0$ in equation
  \eqref{eq:dimensionless-notes} actually shows the ratio
  $E_0/B_0$ by simply regarding $E_0=\Phi_0/r_0$ as a
  typical value for the electric field intensity.
From the idealization of infinite electrical conductivity,
  viz. $\evec +\vvec\times\bvec /c=0$, we know that the
  possible maximum value for the ratio $E_0/B_0 $ is
  approximately $|\vvec|/c$ [n.b. $v_0$ does not denote
  the typical value of $|\vvec|$;
it is a fiducial constant invoked to recover the
  physical dimension of $\vvec$, see definition
  \eqref{eq:fiducial-velocity} in Appendix A].
%   {\bf Meaning?}
From the value of parameter $\epsilon_0$, the definition of
  $v_0$ and Panel B of Fig. \ref{fig:Quadrupole_b_v_field},
  we estimate that $v_0\simeq 1.63\t10t{5}\kps\sim 0.5\ c$
  and $|\vvec|/c \sim 10^{-2}$.
With this value, we have $\delta_0\sim 10^{-3}$ for the
  order of magnitude.
Actually,
% $10^{-3}$ is a sort of maximum value of $\delta_0$;
% {\bf Why?}
  $\delta_0\lsim 10^{-3}$ is also possible under this condition.
   % describing better agreement between magnetic field and flow
   % velocity lines of field.
A smaller $\delta_0$ generally indicates
  that the plasma flow velocity is more
% {\it
  ``closely parallel'' to the magnetic field, or in turn,
  that the electric field is less significant in dynamic
  interactions ($\evec +\vvec\times\bvec/c = 0$).
% }
%   {\bf Physical explanations?}
%  {\bf This statement is not sufficiently accurate.
%  Plasma remains confined to magnetic field lines.
%  I noted this aspect in several places.}

All parameters calculated above are consistent with those
  of Q$_\mathrm{num}$ and Q$_\mathrm{ana}$ MHD models as
  tabulated in Table \ref{table:parameters-model}.
A typical MWD magnetosphere could possibly hold the helical
  magnetized flow configuration, as seen in Panel C of Fig.
  \ref{fig:Quadrupole_b_v_field}, corresponding to a
  rotation of the central MWD.
%{\bf Rotation?}
%{\it The rotation of the central object is surely.}

In addition, MHD models D$_\mathrm{ana}$ and D$_\mathrm{num}$
  can also give sensible dimensional model in the scenario
  of a MWD magnetosphere.
While $\mathcal{M}$, $r_0$, $B_0$ and the dimensionless
  parameter $\gamma_0$ are chosen to be the same as those
  of Q$_\mathrm{ana}$ and Q$_\mathrm{num}$ MHD models,
  parameter $\delta_0$ is set to zero to portray such
  a configuration that
  % {\it
  electric field is negligible, or the magnetic
  field and velocity field more or less coincide.
%  }.
%{\bf Should be more precise here.}
%  {\bf Seems to imply something else! You may also try
%  non-zero $\delta_0$ for rotation and toroidal magnetic
%  field.}{\it already tried for quadrupolar case.}
Dimensionless parameter $\beta_0$,
%  {\bf Related to the usual plasma beta parameter?}
   which is closely related to the well-known
   plasma $\beta$
%  of a plasma
  (n.b. the pressure is somewhat different) is also
  chosen to be zero for the case that magnetospheric
  plasma pressure is completely negligible as
  compared to the magnetic pressure.
The dimensionless parameters
%{\bf What?}
  in Table \ref{table:parameters-model} for the
  D$_\mathrm{ana}$ and D$_\mathrm{num}$ MHD
  models indicate $v_0\simeq 5.16\t10t{4}
  \kps\lsim 0.2\ c$.
  % for $\epsilon_0=10^{-2}$ as presented
  % in Table \ref{table:parameters-model}.

% \subsubsection{Toroidal and Radial Flows of Plasma
%   {\bf How about poloidal?}}

%{\bf How about an isolated neutron star in relatively
% slow rotation?}

%{\it
In general, we can estimate values of the dimensionless
  parameters by specifying the ``typical values'' of some
  important ``fiducial values'' for physical quantities
  in equation \eqref{eq:dimensionless-notes}.
It is fairly straightforward to do so.
  \begin{itemize}
  \item $\beta_0$ by the plasma $\beta$ parameter
    near the surface of the central stellar object;
  \item $B_0$ by the typical magnitude of magnetic
    field near the stellar surface;
  \item $r_0$ by the typical stellar radius;
  \item $\Phi_0/r_0$ by the typical magnitude of electric
    field near the stellar surfac, which can also be
    inferred by $\evec +\vvec\times \bvec/c = 0$;
  \item $\rho_0$ by the typical value of mass density
    near the stellar surface.
  \end{itemize}
For the remaining fiducial parameter $X_0$, which involves the
  toroidal magnetic field, plus the interaction between the
  (poloidal) electric field and the poloidal plasma mass flux.
  Thus $\gamma_0^{1/2}$ may be split into two terms.
The first term is estimated by
\begin{equation}
  \label{eq:gamma-sqrt-toroidal-magnetic}
  \begin{split}
    (\gamma_0^{1/2})_1 & =(1-M_0^2)\left(\dfrac{I_0}{r_0}\right)
    \left(\dfrac{B_0^2}{4\pi} \right)^{-1/2}
    \\
    & = (4\pi)^{1/2}(1-M_0^2)\left( \dfrac{B_{t0}}{B_0}\right)\ ,
  \end{split}
\end{equation}
which compares the typical magnitude of toroidal component ($B_{t0}=
  I_0/r_0$; $I_0$ and $M_0$ are the typical values of $I$ and
  poloidal Alfv\'enic Mach number $M$ near the stellar surface
  respectively) and the total magnetic field ($B_0$).
This term should be close to the order of $(4\pi)^{1/2}\sim 4$
  if we would construct a scenario that the magnetic field
  is dominated by the toroidal component and the poloidal
  Alfv\'enic Mach number $M$ remains small.
The second term should then be estimated by [we also use the second
  and the third lines of equations \eqref{eq:scalar-relation} and
  the definition of poloidal Alfv\'enic Mach number $M$]
\begin{equation}
  \label{eq:gamma-sqrt-toroidal-velocity}
  \begin{split}
    (\gamma_0^{1/2})_2 & = \dfrac{M_0^2}{r_0}\left( I_0 -
      \dfrac{\Theta_0}{F'_0} \right) \left(\dfrac{B_0^2}{4\pi}
    \right)^{-1/2}
    \\
    & = (4\pi)^{1/2} M_0^2\left[ \dfrac{B_{t0}}{B_0} - \left(
    \dfrac{B_{p0}}{B_0}\right)\left(\dfrac{v_{t0}}{v_{p0}}\right)
    \right]\ ,
  \end{split}
  \end{equation}
  which compares $B_{t0}$ with $B_0$, $B_{p0}$ [$B_{p0}=
  (\nabla\psi)_0/r_0$ is the poloidal component of $\bvec$,
  and $(\nabla \psi)_0$ is the typical value of $\nabla\psi$
  near the stellar surface] with $B_0$, and $v_{t0}$ with
  $v_{p0}$ [$\rho v_{t0} = \Theta_0/r_0$ and
  $\rho v_{p0} = (\nabla F)_0/r_0$, where $\Theta_0$
  and $(\nabla F)_0$ are similarly defined as $I_0$
  and $(\nabla\psi)_0$ respectively].

With considerations above, we can discuss (at least to the extent
  of choosing dimensionless parameters) the application of our MHD
  model for isolated neutron stars, which are relatively slow
  rotators (a typical period $P\sim 5\mathrm{\ s}$, or an angular
  speed $\omega_0\sim 1\mathrm{\ rad\ s}^{-1}$) and may not involve
  significant accretions \citep[e.g.][]{2007Ap&SS.308..191V}.
In reference to \citet[][]{1982RvMP...54....1M} and
  \citet[][]{2004ApJ...612.1034P}, we outline an example
  of isolated neutron star as a slow rotator with the
  following parameters:
  $\rho_0\sim 10^{-2}\mathrm{\ g\ cm}^{-3}$ (we take the
  value of $\rho$ at the top of neutron star atmosphere),
  $k_\mathrm{B}T \sim 100\mathrm{\ eV}$ ($k_\mathrm{B}=1.38\times
  10^{-16}\mathrm{\ erg\ K}^{-1}$ is the Boltzmann constant),
  $B_0\sim 10^{13}\gauss$, $r_0\sim 10^6\mathrm{\ cm}$,
  $\mathcal{M} \sim 1 M_\odot$, $\Phi_0/r_0\sim E_0\sim r_0
  \omega_0 B_0/c \sim 10^8\mathrm{\ statV\ cm}^{-1}$.
By selecting these parameters, our further inference may be
  based on the equation of state for an ideal gas ($p\simeq
  \rho k_\mathrm{B} T/m_p$) as well as on
  equations \eqref{eq:dimensionless-notes},
  \eqref{eq:gamma-sqrt-toroidal-magnetic} and
  \eqref{eq:gamma-sqrt-toroidal-velocity}.
The dimensionless parameters are hence estimated as $\beta_0 \sim
  10^{-13}$, $\delta_0\sim 10^{-16}$ and $\epsilon_0\sim 10^{-7}$.
For another parameter $\gamma_0$, we assume that the magnetic field
  is dominated by torodal component, i.e. $B_{t0}/B_0\sim 1$
  and $B_{p0}/B_0\sim 0$, and that the toroidal velocity field
  is not extremely large and hence $v_{t0}/v_{p0}\sim 1$.
These assumptions lead to $\gamma_0\sim 4\pi \sim 10$.
This example shows how the dimensionless MHD model parameters
  can be estimated by sensible physical quantities.
%}

\subsubsection{Outflows of Magnetized Plasmas}
\label{sec:flow}

\begin{figure}
  \centering
  \includegraphics[width=80mm,keepaspectratio]
  {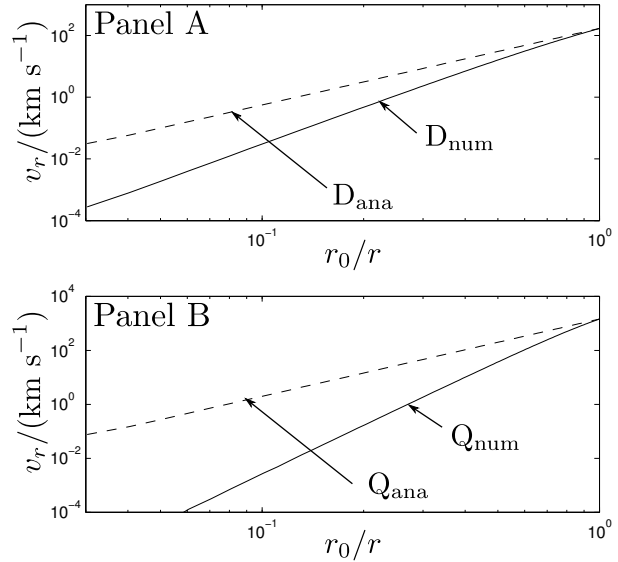}
  \caption{
  Radial flow velocity $v_r$ in unit of $\kps$
    at constant $\mu=-0.9$ versus the inverse
    of radius $r_0/r$ (i.e. $\eta$), indicating
%   an outflow {\bf Breeze? No critical points!}
    an eventual breeze near the polar region.
Such a magnetized breeze does not encounter any MHD critical
    points by the inequality $M^2<1$ enforced to validate
    transformation \eqref{eq:psi-u-transformation}.
  Panel A shows the MHD dipole model (dashed curve for
    D$_\mathrm{ana}$ and solid curve for D$_\mathrm{num}$),
    and Panel B shows the MHD quadrupole model (dashed curve
    for Q$_\mathrm{ana}$ and solid curve for Q$_\mathrm{num}$).
  In both panels, results yielded by semi-analytic models
    are plotted in dashed curves while those by numerical
    models are plotted in solid curves.
  This display is presented in log-log
    scales for a clearer presentation.
  Dimensional recovery is based on the discussion
    in subsection \ref{sec:dimension-recovering}. }
  \label{fig:Outflow}
\end{figure}

\begin{figure}
  \centering
  \includegraphics[width=80mm,keepaspectratio]
  {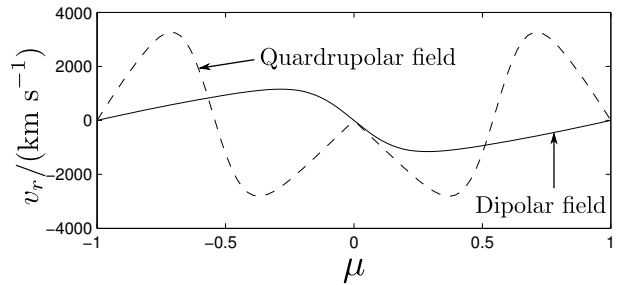}
  \caption{
    The radial velocity for plasma flux at the stellar surface from
      the dipolar models ($D_\mathrm{ana}$ and $D_\mathrm{num}$),
      and the quadrupolar models ($Q_\mathrm{ana}$ and
      $Q_\mathrm{num}$), as functions of $\mu=\cos\theta$.
    Dimensional recovery is based on the discussions
      in subsection \ref{sec:dimension-recovering}
      and Appendix \ref{sec:field-recover}.
    The profiles of dipolar and quadrupolar models are
      presented in solid and dashed curves, respectively.
    }
  \label{fig:Outflow-surface}
\end{figure}

\begin{figure}
  \centering
  \includegraphics[width=80mm,keepaspectratio]
  {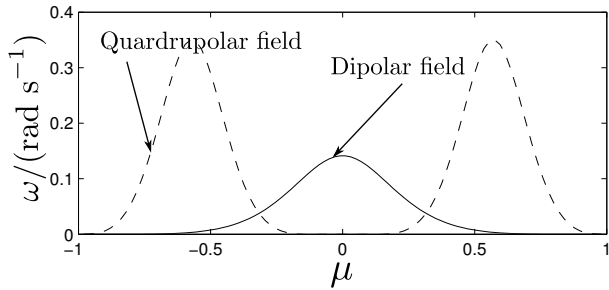}
  \caption{
  %{\bf Type of ``Dipolar" in the figure!!
  % Skype Lile Feb 23, 2013 Saturday}
  Rotation curve over the surface of the central compact
    stellar object, showing the differential angular
    rotation velocity $\omega$ (in radian per second)
    at different $\mu=\cos\theta$.
  This may well describe the case where the compact object
    is covered by a very thin dense ``magnetized plasma ocean''
    \citep[e.g. over the crust of a neutron star as
     described by][]{2001ApJ...563L.147L}.
 %   {\bf Differential rotation? Magnetized plasma
 %   ocean on rotating neutron stars. Lou 2001 on MHD
 %   tidal waves. Extremely strong magnetic fields may
 %   break the solid crust of rotating neutron stars.}
  Dimensional recovery is performed based on the
    discussion in subsection
    \ref{sec:dimension-recovering} and Appendix
    \ref{sec:field-recover}.
%{\bf Please try to construct solutions with
%     solid rotation of compact stellar object numerically.
%     Lile can do so.}
% {\it After thorough re-thinking I guess I cannot construct rigid-body
%   solutions.
%   For a rigid-body, $v_\phi = r_0\omega (1-\mu^2)^{1/2}$,
%   where $(r_0\omega)$ is a constant. If we insert this into equation
%   \eqref{eq:v-field-component}, we will have {\bf INCONSISTENT} results
%   when $\mu\rightarrow \pm 1$. Just multiply the both sides of the
%   third line of \eqref{eq:v-field-component} with $(1-\mu^2)^{-1/2}$,
%   and we will get a constant at LHS. In order to avoid divergence for the
%   first term in the square bracket, we will have to have
%   $u\rightarrow 0$ (to be exact, $(\d f/\d u)x/\varrho$ should goes to zero
%   ``faster'' than $(1-\mu^2)$ if we consider the $b_\phi$ there).
%   However, this implies that the second term in the square bracket
%   should be zero there. But the LHS is a non-zero constant. This
%   requires that the limit $\lim_{\mu\rightarrow 0}[(\d f/\d
%  u)x/\varrho/(1-\mu^2)]$ exists and is non-zero, which is impossible
%   -- we will have a result that is proportional to a positive power
%   of $|u|$ if we use the l'Hospital rule.}
  The results obtained from the dipole models (D$_\mathrm{ana}$ and
    D$_\mathrm{num}$, which give the identical rotation curve) is
    presented by the solid curve, while that from quadrupole model
    (Q$_\mathrm{ana}$ and Q$_\mathrm{num}$, which also give the
    identical rotation curve) is shown by the dashed curve. }
  \label{fig:Diff-rotation}
\end{figure}

As a magnetized gas flows outwards from the vicinity of
  a MWD, an axisymmetric steady flow may be described
  by our MHD model.
We are particularly interested in the radial and toroidal
  components of such MHD flows:
  the radial component gives important information about
  the bulk flow, especially the ``wind" or ``breeze"; the
  toroidal component most clearly shows the interaction
  between the plasma and the magnetic field.
In this subsection, we discuss some profiles of
  flow speed which are of specific interest.
All dimensional quantities can be readily restored in
  reference to subsection \ref{sec:dimension-recovering}.

In Fig. \ref{fig:Outflow}, the radial outflow velocity
  at $\mu=-0.9$ corresponding to a fixed co-latitude
  $\theta$ around a pole is plotted in the logarithmic
  scale for all four MHD model solutions.
We clearly observe that although the numerical model
  solutions are not separable (see subsection
  \ref{sec:not-unique}), they still give radial velocity
  variation in $r$ in an almost power-law profile, just
  similar to that of the semi-analytic models -- yet the
  power indices of the numerical cases are more
  ``negative'' than the semi-analytic cases,
  indicating faster radial falloffs.
This is natural by noting that the $u$ function drops faster
  in the numerical case than in the corresponding
  semi-analytic case, and that $v_r$ is proportional to a
  functional of $u$ as well as the derivative of $u$ with
  respect to $\mu$ [see also equation
  \eqref{eq:v-field-component}].
As we have required $M^2<1$ for poloidal sub-Alfv\'enic
  flows, MHD singularities have been avoided.
With this restriction, we do not have trans-sonic
  and/or trans-Alfv\'enic flows.
%WLL  {\bf Open magnetic field lines?} {\bf No critical points!}
%WLL  {\bf The following replies are not sufficiently clear.
%   July 6, 2013.}
%WLL {\it
%  The light cylinder, approximately, locates at
%    $r_\mathrm{LC}\sim c/\omega_\mathrm{max} \sim 10^{11}\mathrm{\
%      cm}$.
%  This radius corresponds to $\eta\sim 10^{-2}$ ($r_0 = 10^9
%    \mathrm{\ cm}$), where the singularity there could be nicely
%    addressed by neither the semi-analytic nor the numerical scheme
%    (since our models are not relativistic).
%  Fortunately, this singularity will not affect the regions of
%    interest, which are quite close to the central object
%    (e.g. Figures \ref{fig:Dipolar_b_v_field} and
%    \ref{fig:Quadrupole_b_v_field}.)
%WLL  } {\bf Why? July 6, 2013}
%  {\bf Show this more explicitly.}
%{\it I have no idea how to exactly address the power-law
%  profile of snumerical solutions of $u$...}
%In spite of this,
With this in mind, numerical model solutions for $v_r$ fall
  faster than semi-analytic ones do with increasing radius $r$.
For all models, $v_r$ goes to zero as $r$ approaches infinity.
%WLL  {\it
The models are designed to give neither ``winds'' nor ``breezes''
%  {\bf Why not breeze?}
  of MWDs in the global sense; in fact, the global mass loss
  rate from the central stellar object remains zero.
We carefully integrated $v_r$ numerically over the stellar
  surface to find that the overall mass loss rate is virtually
  zero for both dipolar and quadrupolar cases (to be specific,
  the absolute values of mass loss rates are less than
  $\sim 10^{-4}$
%WLL    {\bf what unit?}
  as compared to the total mass exchange rate, which should be
  recognized as nil results considering numerical errors).
Actually, only nils are consistent values for the models.
%WLL  {\bf One may check this from symmetry consideration.}
From equations \eqref{eq:v-field-component}, we have $v_r\propto
  \eta^2(\d f/\d u)(\partial u/\partial \mu)/\varrho$, and thus
  $\lim_{r\rightarrow +\infty}v_r r^2 = \lim_{\eta\rightarrow 0^+}
  v_r/\eta^2 = 0$, if we note that $(\d f/\d u)/\varrho$ must not
  diverge and $\lim_{\eta\rightarrow 0^+}
  (\partial u/\partial\mu) = 0$.
Moreover, we also have $\lim_{r\rightarrow +\infty}\rho v_r r^2 = 0$
  since $\rho\rightarrow 0$ as $r$ goes to (positive) infinity.
Combining those with $\div (\rho\vvec) = 0$ (the steady-state
  mass conservation and the incompressible condition),
%WLL {\bf What?}
  we notice that the total mass flux across any enclosed
  surface in the regions of interest should vanish.
%WLL }
%WLL {\bf Please provide a more complete description for the
%  flow configurations and mass fluxes for all four models. }
%{\it
We also provide a direct illustration of radial flow
  velocity distributrion over the central stellar surface,
  as shown in Fig. \ref{fig:Outflow-surface}.
The quadrupolar models show outflows at higher latitudes and
  inflows near the equator.
For the dipolar field configurations, on the other hand, the
  magnetospheric plasma ``goes out'' from the ``northern''
  hemisphere and ``comes back'' to the ``southern'' hemisphere.
By symmetry properties as stated earlier, our steady MHD models
  remain valid for $\vvec \rightarrow -\vvec$
  and $\evec\rightarrow -\evec$; i.e., we could
  globally reverse the flow velocity field.
Consistent symmetric and topologic features are also
  observed in Fig. \ref{fig:Outflow-surface}.
%} July 6, 2013

If we plot the local angular rotation velocity $\omega$
  versus $\mu$ at $r=r_0$ (i.e. $\eta=\eta_i=1$),
%  {\bf Not at the inner radius?}
  we actually get the latitudinal differential
  rotation profile over the stellar surface.
As displayed in Fig. \ref{fig:Diff-rotation}, the
  rotation curves do not show rigid rotations;
  instead, they show differential rotations.
We expect that this axisymmetric MHD model might be a
    ``portrait'' for astrophysical magnetosphere, whose
    central stellar object has differential rotation over
    the ``surface", e.g. the magnetosphere of a compact
    stellar object or a gaseous planet.
%  {\bf Axisymmetric differential rotation? For your
%  information, the Sun has a surface differential rotation.
%You may want to check the surface differential rotation
% of Jupiter. Jupiter has a magnetosphere.}
For our models, $v_\phi$ vanishes when $u$ goes to zero,
  thus for the two quadrupolar MHD models, Q$_\mathrm{ana}$ and
  Q$_\mathrm{num}$ (they share the same rotation profile since
  $v_\phi$ does not rely on $\partial u/\partial \eta$ or
  $\partial u/\partial\mu$), the surface does not rotate
  at the equatorial plane.
%  {\bf How about other values of $\mu$, away from the equator?}
At higher latitudes, the angular velocity $\omega$ first
  increases then decreases as $\mu$ goes towards $\pm 1$
  for the two poles.
The D$_\mathrm{ana}$ and D$_\mathrm{num}$ MHD models, on
  the other hand, show a more common profile that the
  rotation is fastest near $\mu=0$ at the equator.
%  {\bf Any differential rotation?}
Differential rotation is clearly shown
  in Fig. \ref{fig:Diff-rotation}.
All these MHD models, under the condition of subsection
  \ref{sec:dimension-recovering}, rotates with a period
  of $\sim 10$ to $10^2\mathrm{\ s}$ magnitude as a
  typical value throughout the surface.
This value, in terms of its order of magnitude,
  is consistent with observations
  \citep[e.g.][]{2004ApJ...614..349N}, which gives ``typical''
    rotation periods in the order of tens to hundreds of
    seconds for some fast-rotating white dwarfs.
%  {\bf In what sense? Be more specific.}

%{\bf How about isolated neutron stars?}

\section{Conclusions and Summary}
\label{sec:conclusions-summary}

By adopting and generalizing the semi-analytic formalism
  of \citet{2001GApFD..94..249T} for a specifically
  chosen set of all pertinent free functionals in terms
  of magnetic flux function, we have clearly obtained
  numerically alternative nonlinear steady-state
  axisymmetric MHD solutions with different
  characteristics and features in reference to the
  corresponding separable semi-analytic nonlinear
  solutions with the same boundary conditions.
The governing MHD nonlinear PDE is a quasi-linear
  elliptic equation, which would possess a unique
  solution when the corresponding functional
  properly defined has only one extremum.
As demonstrated and elaborated in subsection
  \ref{sec:sufficient-cond-unique}, the above
  sufficient condition for the uniqueness of nonlinear
  solutions is actually not met for the chosen free
  functionals of \citet{2001GApFD..94..249T}.
This merely opens up the possibility for multiple
  nonlinear solutions with the same boundary
  conditions but not necessarily so.
Our extensive numerical calculations with FDM and FEM codes
  clearly show that those semi-analytic separable solutions
  are indeed not unique, even though all boundary conditions
  (at the stellar surface and infinity) and flux functionals
  are chosen to be identical with those of semi-analytic
  2.5D steady MHD models.
%{\bf What are the specific radial fall-offs
%  of the magnetic field strengths for numerical
%  solutions?}
%{\it
The numerical solutions have the same topological structures when
  compared with their corresponding separable semi-analytic solutions,
  yet one of the most salient features of the numerical solutions is
  that the magnitudes of the magnetic and plasma flow velocity fields
  fall faster than those of semi-analytic solutions do, respectively.
% }
%{\it
An extremely important implication of non-uniqueness of
  solutions due to nonlinearity is that the extrapolation
  codes, which are mostly used for rebuilding three-dimensional
  configurations of solar/stellar or astrophysical disk coronal
  magnetic fields from the photospheric boundary conditions,
  may not work well for nonlinear cases of steady MHD and/or
  magnetostatic equilibria.
One actually needs additional information and requirements
  in order to determine possible field configurations.
Pertinent discussions, which are also related to
  \citet{1993A&A...278..589B}, are included in
  Appendix \ref{sec:prev-works-multi}
%  }

%  {\bf Please highlight the consequences for
%  numerical scheme of extrapolation to infer
%  solar/stellar coronal magnetic fields.
%Please read Low \& Lou (1990) (and several
%  references therein) carefully regarding
%  ill-posed problems in particular.
%Note that the solution to a potential field problem
%  ($\alpha=0$ in the force-free condition) is unique
%  -- the discussion and arguments of Low \& Lou
%  (1990) is based on that.
%For a constant $\alpha$, the force-free problem
%  is linear and the solutions are unique.
%The non-uniqueness and/or multiple solutions to
%  the force-free problem and our more general
%  steady MHD problem here pose further challenges
%  to numerical extrapolation schemes. }

We further explore
  our numerical MHD models to compare with the semi-analytic ones
  and find that magnitudes of important quantities (such as
  dimensionless $\bvec/B_0$, $\vvec/v_0$, etc.) of numerical
  models fall faster as radius $r$ goes to infinity.
This reflects that the energy of the separable semi-analytic
  MHD models are higher than their corresponding
  numerical counterparts.
This perspective is specifically verified by calculating
  and comparing the pertinent magnetic and kinetic
  energies numerically.
The difference in energies implies that the numerical
  solutions are more stable when other conditions remain
  the same, which might explain the fact that relaxation
  methods always converge to a solution different from
  the corresponding semi-analytic ones when the outer
  radius is sufficiently large (i.e. $\eta_o\rightarrow 0^+$).
Our numerical explorations open up the possibility
  of constructing even more solutions in addition
  to what we have found so far.
%{\it However, we still claim that it is possible that the numerical
%  results does not indicate a minimum of energy [although it has to
%  be a extremum of the functional \eqref{eq:u-pde-functional}]. }
%  {\bf Please think more on this.}
The situation is similar and/or parallel to what was
  constructed in \citet{1993A&A...278..589B} for a 2.5D
  static force-free magnetic field configuration (see
  Appendix \ref{sec:prev-works-multi} for more details).

We propose that this class of 2.5D steady MHD model
  solutions may be
  applicable for conceiving astrophysical scenarios for
  magnetospheres of MWDs with surface differential
  rotations at the photospheres given a proper choice
  of several specific parameters (e.g. the dimensionless
  parameters $\beta_0$, $\gamma_0$, $\delta_0$ and
  $\epsilon_0$, the fiducial magnitudes of magnetic field
  and flow velocity field), which requires input of
  pertinent data, calculations as well as physical estimates.
Conceptually, it is also possible to extend our 2.5D steady
  MHD model considerations to neutron star magnetospheres
  if we ignore general and special relativistic effects as
  simplifications.
Regarding the surface differential rotation of neutron
  stars, it was suggested by Lou (2001) that there may
  exist a very thin, dense, and intensely magnetized
  plasma ocean over the surface of a neutron star (i.e.
  the base of a neutron star magnetosphere).
Magnetic field models with dipolar and quadrupolar
  configurations can be obtained systematically --
  they generated from semi-analytic and numerical
  solution models, respectively, by consistently
  choosing pertinent parameters.
We note in particular that under certain conditions
  (for example, when the lines of flow velocity
  field and of the magnetic field do not coincide
  with each other in the presence of considerable
  electric field), there can be helical MHD outflows
  and/or inflows around the polar regions of a
  stellar magnetosphere.
%   {\bf Related to rotation!}
Physically, we attribute this phenomenon to the
  surface differential rotation of the central star.
By more extensive and detailed calculations, we find
  that the radial outflow velocity approaches zero at
  infinite radius, which is a generic feature for a
  ``magnetized stellar breeze''.
%  {\bf Please compute the mass loss rates in the current
%  context of MWDs.}
%{\it
The global net mass loss rate remains zero over the
  spherical stellar surface, which is consistent with
  the assumption of steady state and the vanishing
  radial velocity at infinite radius.
%  }
%{\bf Need to know the overall flow configurations!}
%{\it
The net mass flow across any enclosed surface is proven
  to be zero, and the radial flow features around the central
  stellar object shows conserved mass flows, which are spewed
  into the magnetosphere, travel around, and then flow back
  to the stellar surface.
%  }
For applications,
%{\it
  the values of the dimensionless parameters can also
  be estimated in reference to the ``typical values'' of
  astrophysical quantities around the central stellar object.
%  }
%We specified those
Our 2.5D steady MHD models indicate differential rotations
  at the compact stellar surface, giving a typical rotation
  periods consistent with observations, which is about
  $10\sim 10^2$ seconds.
%  {\bf Provide specific numbers.}

\section*{Acknowledgments}

This research was supported in part by Ministry of Science
 and Technology (MOST) under the State Key Development
 Programme for Basic Research grant 2012CB821800,
 %  射电波段的前沿天体物理课题及FAST早期科学研究 公示内容
 % This work is partly supported by China Ministry of
 % Science and Technology under State Key Development
 % Program for Basic Research (2012CB821800).
 by the Tsinghua University Initiative Scientific Research Programme
 (20111081008), by the Special Endowment for Tsinghua College
 Talent (Tsinghua XueTang) Programme from the Ministry of
 Education (MoE),
 by the National Natural Science Foundation of China (NSFC)
 grants 10373009, 10533020, 11073014 and J0630317 at Tsinghua
 University, by the Tsinghua Centre for Astrophysics (THCA),
 and by the SRFDP 20050003088, 200800030071 and 20110002110008,
 and the Yangtze Endowment from the MoE at Tsinghua University.

% \vskip 0.5cm
\vspace{0.5cm}

%{\bf Several references are
% indicated in the maintext! Please add them
% into the reference list below. }

\bibliographystyle{mn2e}
\bibliography{MHD3R}

% {\bf Lou 2001 on MHD tidal waves added.}
% {\bf Lou 1995 on MWD oscillations added.}

\appendix

\section{From Reduced Scalar Nonlinear MHD PDE
  to Dimensionless Magnetic and Velocity Fields}
\label{sec:field-recover}

In principle, the magnetic field can be derived
  from the $u(\eta,\ \mu)$ solution of the reduced
  nonlinear MHD PDE \eqref{eq:u-pde-dimensionless}
%  (i.e. the $u$ function)
  by properly combining equations
  \eqref{eq:vect-quant}, \eqref{eq:scalar-relation},
  \eqref{eq:psi-u-transformation} and
  \eqref{eq:power-func-u}.
However, specific details are not as trivial as
  thus simply stated.

By inspecting equation \eqref{eq:vect-quant},
  we here define two more dimensionless quantities,
  viz. the dimensionless magnetic field $\mathbf{b}$
  and $f$ -- the dimensionless form of the free
  functional $F$; they are respectively
\begin{equation}
  \mathbf{b} = \bvec/B_0\ ,\qquad\qquad
  F(U) = U_0\rho_0^{1/2}f_0f(u)\ ,
\end{equation}
where $f_0$ is a dimensionless parameter
  introduced to simplify the definition of $f(u)$.

By equations \eqref{eq:dimensionless-trans} and
  \eqref{eq:dimensionless-notes}, we thus obtain the
  expressions for the three components of $\mathbf{b}$
  field (n.b. here $x$ is a functional of $\psi$ as
  introduced in eq. \eqref{eq:dimensionless-trans},
  not a Cartesian coordinate component):
\begin{equation}
  \label{eq:b-field-component}
  \begin{split}
    b_r & = \dfrac{\eta^2}{\alpha} (1-M^2)^{-1/2}
    \dfrac{\partial u}
    {\partial \mu}\ ,
    \\
    b_\theta & = -\dfrac{\eta^3}{\alpha} (1-M^2)^{-1/2}
    (1-\mu^2)^{-1/2} \dfrac{\partial u} {\partial \eta}\ ,
    \\
    b_\phi & = \eta (1-\mu^2)^{-1/2} \bigg[ \gamma_0^{1/2}
    (1-M^2)^{-1/2} x
    \\
    &\qquad\qquad\qquad\qquad
    - \delta_0^{1/2}f_0 \left( \dfrac{1-\mu^2} {\eta^2} \right)
    \dfrac{\d f}{\d u} \dfrac{\d \varphi} {\d u} \bigg]\ .
  \end{split}
\end{equation}
After obtaining three component expressions of
  dimensionless $\mathbf{b}$ field in equation
  \eqref{eq:b-field-component}, we would express
  the square of poloidal Alfv\'enic Mach number
  $M^2$ as $(F')^2/\rho$, or further in terms of
  $\d f/\d u$.
By applying integral transformation
  \eqref{eq:psi-u-transformation}, we readily
  have $\d U = (1-M^2)^{1/2} \d \psi$ and thus
\begin{equation}
  \label{eq:df_dpsi-to-df_du}
  \begin{split}
    M^2 & = \dfrac{(F')^2}{\rho} =
      \dfrac{(1-M^2)}{\rho} \left(
      \dfrac{\d F} {\d U} \right)^2
    \\
    & = \dfrac{f_0^2(1-M^2)} {\varrho} \left( \dfrac{\d f}{\d u}
    \right)^2\ .
  \end{split}
\end{equation}
From this, we readily derive an expression for
  $(1-M^2)$ in terms of $\d f/\d u$ and $\varrho$, viz.
\begin{equation}
  \label{eq:m-in-df_du-varrho}
  1-M^2 = \left[ 1 + \dfrac{f_0^2}{\varrho} \left(
      \dfrac{\d f} {\d u} \right)^2 \right]^{-1}\ .
\end{equation}
By equation \eqref{eq:m-in-df_du-varrho},
  the poloidal sub-Alfv\'enic MHD flow condition
  of $M^2<1$ clearly remains always satisfied
  in our MHD model formulation.

Parallel procedures are taken for the
  plasma flow velocity field.
First we define the fiducial magnitude of
  flow velocity $v_0$ as
\begin{equation}
  \label{eq:fiducial-velocity}
  v_0 = \left( \dfrac{2G\mathcal{M}}{\epsilon_0 r_0}
  \right)^{1/2}\ .
\end{equation}
% {\bf This $v_0$ parameter appeared in the main
% text but was not defined there somehow!}
%{\it addressed in the main text now.}
With this definition \eqref{eq:fiducial-velocity}
  and equation \eqref{eq:m-in-df_du-varrho}, the
  dimensionless expressions of the flow velocity
  components are
%  can be obtained by the following expressions:
\begin{equation}
  \label{eq:v-field-component}
  \begin{split}
    & \dfrac{v_r}{v_0} = \dfrac{\eta^2f_0}{\alpha\varrho}
    \dfrac{\d f}{\d u} \dfrac{\partial u} {\partial\mu}\ ,
    \\
    & \dfrac{v_\theta}{v_0} = -\dfrac{\eta^3f_0}{\alpha\varrho}
    (1-\mu^2)^{-1/2} \dfrac{\d f} {\d u} \dfrac{\partial u}
    {\partial\eta}\ ,
    \\
    & \dfrac{v_\phi}{v_0} = (1-M^2)^{1/2} \left[ \dfrac{b_\phi
        f_0}{\varrho} \dfrac{\d f}{\d u} - \delta_0^{1/2}
      \dfrac{(1-\mu^2)^{1/2}}{\eta} \dfrac{\d\varphi}{\d u}
    \right]\ .
  \end{split}
\end{equation}
%{\bf Please make sure carefully that these expressions
% as well as others are free of errors and typos!}
%
Note that we have not yet specified the form
  of $F(U)$, or $f_0$ and $f(u)$ -- neither
  did \citet{2001GApFD..94..249T},
%  {\it
  as it is not directly involved in reduced
  nonlinear PDE \eqref{eq:u-pde-dimensional}.
%  }
%{\it
The free functional $F$ actually gives another
  degree of freedom for ``constructing'' the plasma
  flow velocity field from the solution of nonlinear
  reduced PDE \eqref{eq:u-pde-dimensional}.
While a solution to PDE \eqref{eq:u-pde-dimensional}
  is obtained, a different definiton of $f(u)$ will
  actually give rise to a different configuration
  of the flow velocity field.
In order to provide specific examples in
  the main text for elaboration, we prescribe
  an $f(u)$ right here.
% }
%{\bf Why not making this specification earlier
% in the main text together with equation
% (9)? Not necessarily involved?}
An expedient choice is adopted to simply set $f_0=10^{-2}$
  in the non-relativistic regime and require
  $(\d f/\d u)/\varrho$ not to diverge as
  $u\rightarrow 0$, e.g.
\begin{equation}
  \label{eq:f-function-form}
  f= \dfrac{\alpha}{(4\alpha+3)} |u|^{4+3/\alpha}\ .
\end{equation}
Throughout Sections \ref{sec:numerical-results}
  and \ref{sec:applicatoins}, we have actually used
  this functional expression \eqref{eq:f-function-form}
  as well as those prescribed in equation
  \eqref{eq:power-func-u}.
Based on these ansatz, remaining steps of getting
  dimensionless magnetic field and plasma flow
  velocity fields then become fairly straightforward.

\section{On multiple solutions to nonlinear
   force-free magnetic field equations}
\label{sec:prev-works-multi}

\begin{figure*}
  \centering
  \includegraphics[width=160mm,keepaspectratio]
  {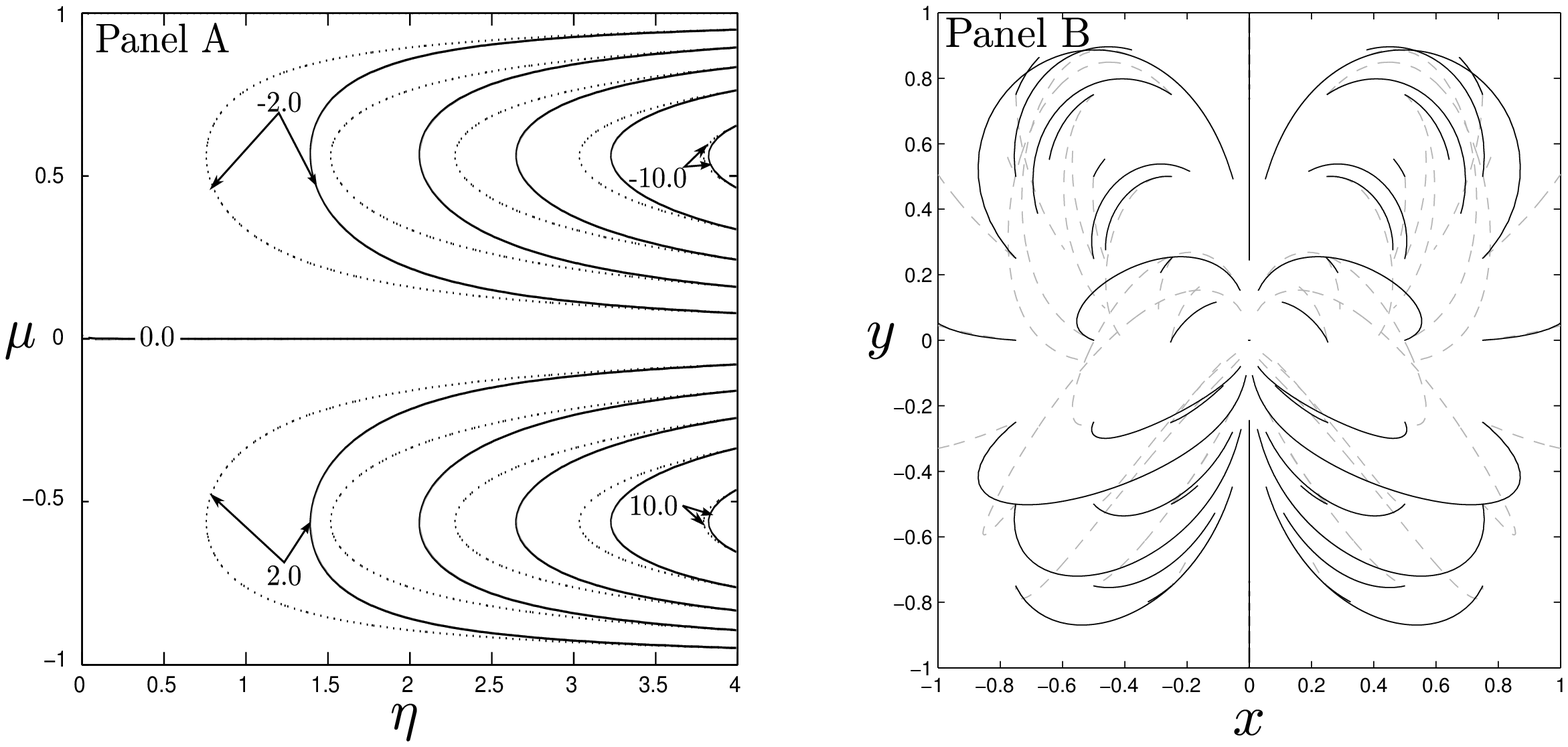}
  \caption{
    Comparison between the numerical solution and the corresponding
      separable semi-analytic solution with $n=m=1$ (i.e. the
      eigenvalue $a^2_{1,1}=0.4274$) in the problem of force-free
      magnetic field as studied by \citet{1990ApJ...352..343L}.
    Panel A shows a contour plot, comparing the
      separable semi-analytic solution (dotted contour curves)
      with the numerical one (solid contour curves).
    Adjacent contour curves are separated by $2$, while
      some fiducial contour curves are labelled.
    Panel B shows magnetic field lines generated from
      the semi-analytic solution (dashed curves) and
      the numerical solution (solid curves).
    Those magnetic field lines are generated by the
      same scheme as in panel (b) of figure 3 in
      \citet{1990ApJ...352..343L}.
    The differences are conspicuous. }
  \label{fig:Example_previous_work}
\end{figure*}

%WLL   {\bf As discussed on April 5-6, 2013, please
% discuss and elaborate possible consequences
% of multiples nonlinear solutions to the
% force-free magnetic field problem in the
% main text and here when appropriate.
%The key issue is related to those
% extrapolation numerical
% simulations (please include pertinent
% references) using the semi-analytical
% solutions of Low \& Lou (1990) as
% the testing reference.
%Please make clear statements of our
% point of view.}

Our work here is in fact inspired by our previous
  completely independent research work, which
  turned out to reproduce the results of
  \citet{1993A&A...278..589B} by alternative
  numerical methods.
In Appendix B here, we summarize the key pertinent
  results and point out possible consequences on
  numerical simulations to extrapolate solar/stellar
  and/or astrophysical disk coronal magnetic field
  configurations by specifying boundary conditions
  at a given surface (e.g. photosphere).

First, we present the fundamental MHD equations for
 the so-called force-free magnetic field condition
  \citep[e.g.][]{1958PNAS...44..285C}:
\begin{equation}
  \label{eq:force-free-vector-pde}
  \curl \bvec =\bar \alpha \bvec\ ,\qquad\qquad\qquad
  \bvec \cdot \nabla\bar \alpha = 0\ ,
\end{equation}
where the divergence-free condition for magnetic field
 $\bvec$ is imposed.
Here $\bar \alpha$ is a scalar field describing the ratio between
  the magnetic field and the electric current density vector
%  the velocity field
%  {\bf Electric current density vector! Mar 30, 2013!}
  (n.b. these two vector fields are parallel to each other
  everywhere under the force-free condition).
According to \citet{1990ApJ...352..343L}, force-free magnetic
  field equation with axisymmetry can be recast into the
  following form (here we still adopt $\eta = r_0/r$ and
  $\mu =\cos\theta$):
\begin{equation}
  \label{eq:force-free-scalar-pde}
  \dfrac{\partial }{\partial \eta} \left( \eta^2\dfrac{\partial
      A}{\partial \eta} \right) + (1-\mu^2) \dfrac{\partial^2 A}
  {\partial \mu^2} + \dfrac{Q}{\eta^2} \dfrac{\d Q}{\d A} =
  0\ ,
\end{equation}
%{\bf Typo in the second term on the left-hand side!
% Should be $\mu$ instead of $\eta$!}
where $A(\eta,\ \mu)$ is a scalar function giving the magnetic
  field $\bvec$ in the following form with $\rvec$, $\thvec$
  and $\phivec$ being orthogonal unit vectors for spherical
  polar coordinates ($r,\ \theta,\ \phi$):
\begin{equation}
  \label{eq:force-free-vector-field}
  \bvec = \dfrac{1}{r\sin \theta} \left( \dfrac{1}{r}
    \dfrac{\partial A}{\partial \theta} \rvec -
    \dfrac{\partial A}{\partial r} \thvec + Q\phivec
  \right)\ ,
\end{equation}
%{\bf Missing $r$ in the denominator!}
and $Q = Q(A)$ is a
  free functional of $A$ related to the
  toroidal component of the magnetic field.
%  {\bf Please indicate $\alpha$-$Q$ relation!}
%{\it We note here that
Note that $\bar \alpha$ is related to $Q$ by
  the relation $\bar \alpha = \d Q / \d A$.
\citet{1990ApJ...352..343L} selected a specific
  pair of $A$ and $Q(A)$ such that separable
  nonlinear solutions can be achieved
\begin{equation}
  \label{eq:force-free-separable}
  A = \dfrac{P(\mu)}{r^n}\ ,\qquad\qquad\qquad
  Q(A) = a A |A|^{1/n}\ ,
\end{equation}
where $n$ is an index and $a$ is an
  arbitrary real parameter.
This scheme yields a separable
 nonlinear ODE of $P(\mu)$, viz.
\begin{equation}
  \label{eq:force-free-separated-ode}
  (1-\mu^2) \dfrac{\d^2 P}{\d \mu^2} + n(n+1) P
  + a^2\dfrac{(1+n)}{n} P^{1+2/n} = 0\ .
\end{equation}
%{\bf Typo again for 2.}
Proper boundary conditions for this nonlinear
  ODE, i.e. $P_{\mu=\pm 1} = 0$,
  should be imposed such that magnetic fields
  $\bvec$ at the two poles $\mu=\pm 1$ do not diverge.
These boundary conditions only allow nonlinear
  ``eigensolutions'' to exist, corresponding
  to discrete different $a^2$ as ``eigenvalues''.
We denote different nonlinear eigensolutions by
  the rising sequence of their eigenvalues,
  i.e. the $m$th eigenvalue with a given $n$
  is indicated by $a^2_{n,m}$, as similarly
  done in \citet{1990ApJ...352..343L}.

%{\bf Stop here!}

% Readers may notice here that equation
%   \eqref{eq:force-free-separated-ode} is very similar to
%   equation \eqref{eq:u-h-separated-ode} in case of $\beta_0
%   = \gamma_0 = \delta_0 = \epsilon_0 = 0$.
% However, we can tell the difference since $\epsilon_0=0$ is
%   unacceptable (see Appendix \ref{sec:field-recover}).

After determining eigenvalues and eigensolutions numerically
  from nonlinear ODE \eqref{eq:force-free-separated-ode}, a
  series of discrete solutions with the following specific
  boundary conditions
\begin{equation}
  \label{eq:force-free-bound-cond}
  A|_{\eta=\eta_0} = P(\mu)\eta_0^n\ ,\quad\qquad
  A|_{\eta\rightarrow 0^+} = A|_{\mu=\pm 1} = 0\ ,
\end{equation}
  is therefore obtained by $A=P(\mu)\eta^n$ and further
  information for the toroidal magnetic field is given
  by $Q(A) = (a_{n,m}^2)^{1/2} A|A|^{1/n}$.
For a rescaling of $P(\mu)$ to $P_0P(\mu)$ with $P_0$
  being a free normalization, we must require $a_{n,m}^2$
  to $a_{n,m}^2P_0^{2/n}$ accordingly.
We have already remarked in the main text that eigensolutions
  of nonlinear ODE \eqref{eq:u-h-separated-ode} for $h(\mu)$
  no longer possess such rescaling property because of the
  summation of several power-law terms on the RHS.
In other words, for a similar rescaling of $h(\mu)$, one can
  readily determine the correspondingly shifted eigenvalues.
Nevertheless, the free normalization and the corresponding
  eigenvalues are not governed by a simple analytic relation
  as shown above for the force-free case.
Basically, we have to compute them on the one-to-one basis.

%WLL  {\bf Regarding the rescale property of the force-free
%magnetic field problem, it is worthwhile to construct
%numerical solutions corresponding to several rescaled
%yet equivalent semi-analytical solutions.
%  It is important to check and compare resulting
%  numerical solutions in the main text and/or here.
%  I am curious whether the resulting numerical
%  solutions remain rescalable in a similar manner.  }
\begin{figure}
  \centering
  \includegraphics[width=70mm,keepaspectratio,trim=10 10 10 10]
  {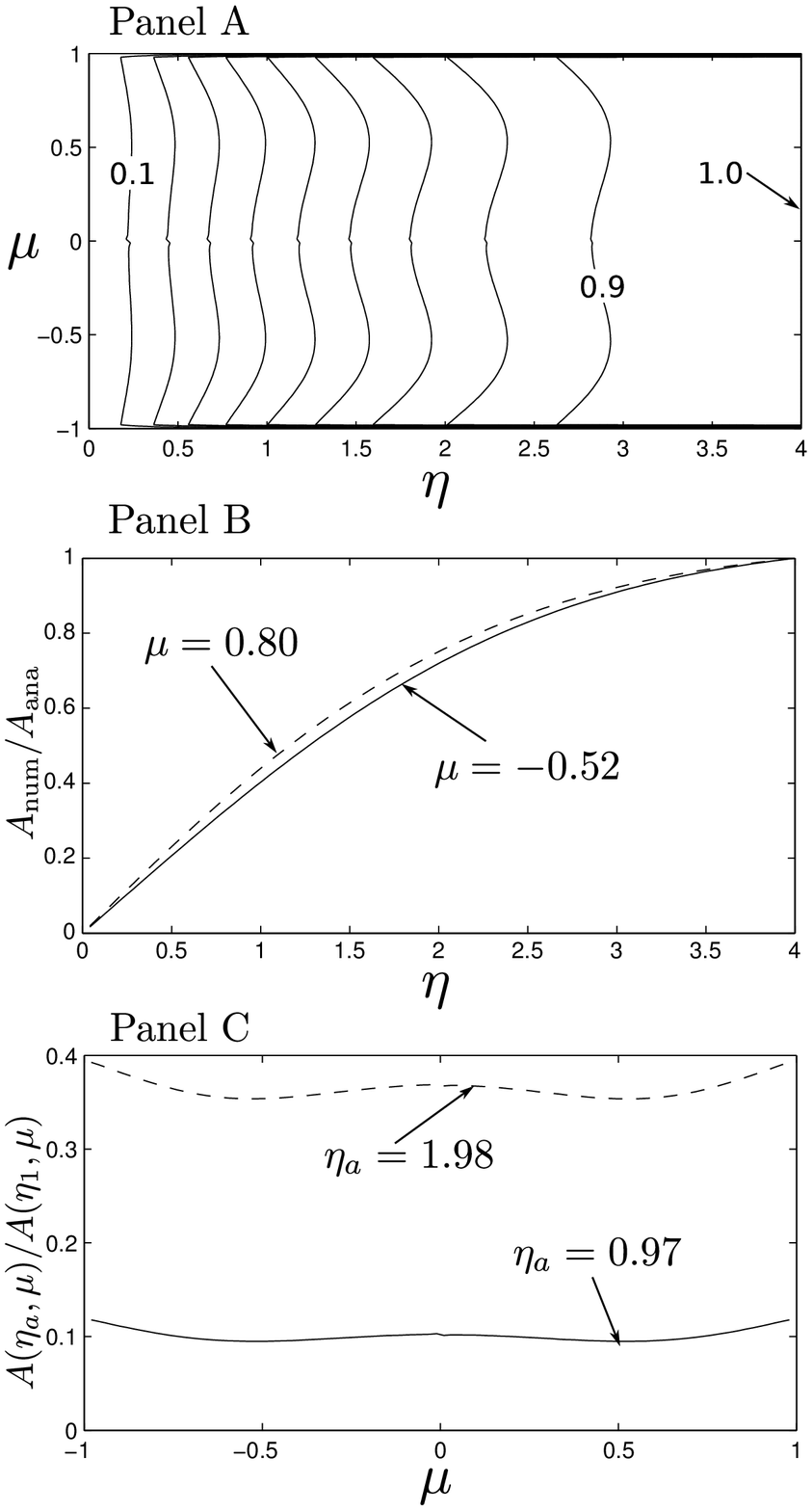}
  \caption{
    Here, we compare our new solution to the semi-analytic
      solution in \citet{1990ApJ...352..343L}.
    We further show that the new solution is not separable.
    Panel A presents the contour plot of$A_\text{num}/A_\text{ana}$
    in the complete spatial domain where the nonlinear
    PDE is solved (difference of elevation
    between adjacent contours is $0.1$).
    Panel B shows $A_\text{num}/A_\text{ana}$ when $\mu$
    are assigned two different values, viz. $\mu=0.80$
    (dashed curve) and $\mu=-0.52$ (solid curve).
    Panel C shows that the new solution is not a separable solution:
      $A(\eta_a,\mu)/A(\eta_1,\mu)$ are presented with fixed $\eta_a$
      ($\eta_1=4$ is  the value of $\eta$ at the boundary).
    We have taken $\eta_a=1.98$ for the dashed curve and
      $\eta_a=0.97$ for the solid curve.}
    \label{fig:Previous-non-rescale}
\end{figure}

%WLL  {\bf The following added paragraph (checked and by themselves
% fine) missed the point of rescale though. July 8, 2013}

%{\it
We compare in Figure \ref{fig:Previous-non-rescale}
  the semi-analytic solution with the new solution
  (subscript ``num'' denotes the new numerical solution).
We also show in this figure that the new numerical solution
  is {\it not} a separable solution.
Panel A reveals that $A_\text{num}/A_\text{ana}$ varies
  significantly, while a result of rescaling would give
  a constant ratio.
Panel B, which shows $A_\text{num}/A_\text{ana}$ with a fixed
  $\mu$ value, illustrates that the new numerical solution descends faster
  to zero than the semi-analytic solution when $r$ goes to $+\infty$
  (or equivalently, when $\eta$ goes to $0^+$).
It also tells the impossibility of separating variables: they indicate
  that $A_\text{num}/A_\text{ana}$ has different patterns if we assign
  $\mu$ to different values and plot $A_\text{num}/A_\text{ana}$
  against $\eta$.
Panel C shows the non-separability in another aspect: we present the
  variation of $A(\eta_a,\mu)/A(\eta_1,\mu)$, where $\eta_a$ is a
  constant $\eta$ value and $\eta_1$ is value of the $\eta$ at the
  boundary.
Clearly $A(\eta_a,\mu)/A(\eta_1,\mu)$ is not a constant, which is in
  contrast with separability.
%}

%{\bf Stop here!}

Nevertheless, as was found in our extensive numerical explorations
  using both FDM and FEM schemes, when we prescribe the boundary
  condition \eqref{eq:force-free-bound-cond} (with $\eta_0=4$ as
  an example)
%WLL   {\bf Need a check of notation and meaning!
%  Also invoked presently in this appendix. }
  for nonlinear PDE \eqref{eq:force-free-scalar-pde},
  we would obtain via iterations a numerical solution
  which is definitely different from the separable
  semi-analytic nonlinear solution.
At least, this can be done in a sequence of nonlinear
  solution pairs with the same boundary conditions.
As an example of illustration, we present in Panel A of
  Figure \ref{fig:Example_previous_work} a comparison
  between a pair of semi-analytic results with $n=1$,
  $m=1$ (i.e. $a^2 = a^2_{1,1}= 0.4274$), and its
  corresponding numerical nonlinear solution
  determined by the above procedure.
Meanwhile in Panel B of Figure
  \ref{fig:Example_previous_work}, we
  compare magnetic field lines of force
  generated by two different solutions in the pair.
For each type of solutions, we plot 49 magnetic field lines
  of force starting from 49 evenly spaced points lying
  on the plane with $z=0.3$ and $-1<x,\ y<1$ (these are
  Cartesian coordinates), as was done in panel (b) of
  figure 3 by \citet{1990ApJ...352..343L}.
The differences are patently clear. The existence of
 multiple nonlinear solutions prompt us to further
 explore the steady MHD problem with possible multiple
 nonlinear solutions described in this paper.

We highlight the following statements regarding
 the development and tests of numerical
 extrapolation MHD codes over decades to construct
 various configurations of solar/stellar and
 astrophysical disk coronal magnetic fields.
For example in the solar case, it is highly desirable for
 solar physicists to infer coronal magnetic fields based
 on the measurements of solar photospheric magnetic fields
 \citep[e.g.][and references therein]{1990ApJ...352..343L}.
% (e.g. Low \& Lou 1990 and references therein).
Realizing the real possibility of multiple
 (currently at least two by our MHD model
 analysis) nonlinear solutions with the
 exactly same boundary conditions, the
 numerical code development for extrapolating
 magnetic field configurations from specified
 or measured surface conditions would face considerable
 challenges and should proceed with extreme cares.
Essentially, this dilemma would involve computation
 selection effects of numerical solutions and nature
 selection effects of physical solutions.
Two key questions are then:
 How do we know that our extrapolation MHD codes are
 reliable through tests against known solutions?
 What are the paths of choosing separate
 numerical solutions?
In a sense, we need to know and determine all
 multiple solutions for the same boundary conditions
 if there are indeed more than two.
Otherwise, we would be highly uncertain about the
 convergence of our numerical solutions as settled
 by our extrapolation MHD codes.
One conceivable yet unfavorable situation would be
 the undesirable artificial exclusion of possible
 numerical solutions because of the implementation
 of a particular numerical scheme under certain
 constraints.
Even if all numerical solutions with the same
 boundary conditions are know and available,
 we would then need additional physical
 information and/or requirements to remove
 such degeneracy of multiple nonlinear solutions.
Or, they are all realizable in nature.
We would like to know their stabilities as
 well as their mutual or independent evolution.
This appears to be a new dimension/perspective
 for this research field involving nonlinear
 degeneracy.
 %(\bf Added July 14, 2013}

%  {\bf Implications to numerical simulations
%  of solar/stellar coronal magnetic field
%  configurations should be discussed clearly
%  in the main text.}
% {\bf I also mentioned this in the main text
%  regarding ill-posed problem (Low \& Lou
%  1990 and references therein and multiple
%  solutions (non-uniqueness). }
%{\it
One implication of these findings for multiple
  (i.e. non-unique) nonlinear solutions is that they might
  cause ``peculiar things'' that may happen in simulation
  tests for numerical MHD codes to extrapolate solar/stellar
  and astrophysical disk coronal magnetic fields from the
  photosphere, especially those likely involving aspects
  of the ``ill-posed'' problems as emphasized by
  \citet{1990ApJ...352..343L}.

For a clearer presentation of this problem and further
  discussion, we indicate below four common
  ``diagnostics'' \citep[e.g.][] {2006A&A...446..691A}
  of our numerical nonlinear solutions, one of which is fairly
  similar to equation \eqref{eq:c-s-correlation} in subsection
  \ref{sec:dimension-recovering}.
First, we state the mathematical definitions of
  those ``diagnostic functions''. The correlation
  of vector magnetic fields:
%  {\bf Stop here!}
\begin{equation}
  \label{eq:vec-corr}
  \text{Corr}_\text{v} ( \mathbf{B}_\text{ana}, \mathbf{B}_\text{num} )
  = \dfrac{\int_V \mathbf{B}_\text{ana}\cdot \mathbf{B}_\text{num}
    \text{d} \tau } {\left(\int_V |\mathbf{B}_\text{ana}|^2
    \text{d} \tau \right)^{1/2} \left(\int_V |\mathbf{B}_\text{num}|^2
    \text{d} \tau \right)^{1/2}}\ ;
\end{equation}
%  {\bf Define notations $dV$ and $d\tau$!}
  the Cauchy-Schwartz correlation of two vector magnetic fields:
\begin{equation}
  \label{eq:cs-corr}
  \text{Corr}_\text{cs}( \mathbf{B}_\text{ana}, \mathbf{B}_\text{num} )
  = \dfrac{1}{V} \int_V \dfrac { \mathbf{B}_\text{ana}\cdot
    \mathbf{B}_\text{num} }
  {| \mathbf{B}_\text{ana} | | \mathbf{B}_\text{num} | } \text{d} \tau
  \ ;
\end{equation}
  the normalized error of a vector magnetic field:
\begin{equation}
  \label{eq:norm-vec-err}
  \text{Err}_\text{n} ( \mathbf{B}_\text{ana}, \mathbf{B}_\text{num} )
  = \dfrac{\int_V | \mathbf{B}_\text{ana} - \mathbf{B}_\text{num}|
    \text{d} \tau } {\int_V |\mathbf{B}_\text{ana}| \text{d} \tau }\ ;
\end{equation}
  and the mean error of a vector magnetic field:
\begin{equation}
  \label{eq:mean-vec-err}
  \text{Err}_\text{m} ( \mathbf{B}_\text{ana}, \mathbf{B}_\text{num} )
  = \dfrac{1}{V} \int_V \dfrac { | \mathbf{B}_\text{ana} -
    \mathbf{B}_\text{num} | } { | \mathbf{B}_\text{num} | } \text{d} \tau \ .
\end{equation}
In the foregoing four expressions, $\text{d}\tau$ is the differential
  volume element and $V$ is the spatial volume for integration.
Clearly, if the two vector fields are identical everywhere inside
  the spatial volume $V$ for integration, we should have
  $\text{Corr}_\text{v} =
  \text{Corr}_\text{cs} = 1$ and $\text{Err}_\text{n} =
  \text{Err}_\text{m} = 0$.
In other words, the amounts by which the correlation functions
  ($\text{Corr}_\text{v}$ and $\text{Corr}_\text{cs}$) deviate
  from $1$ and/or the error functions ($\text{Err}_\text{n}$ and
  $\text{Err}_\text{m}$) deviate from $0$ indicate the extent for
  which $\mathbf{B}_\text{num}$ deviates from $\mathbf{B}_\text{ana}$.

\begin{figure}
  \centering
  \includegraphics[width=70mm,keepaspectratio,trim=10 10 10 10]
  {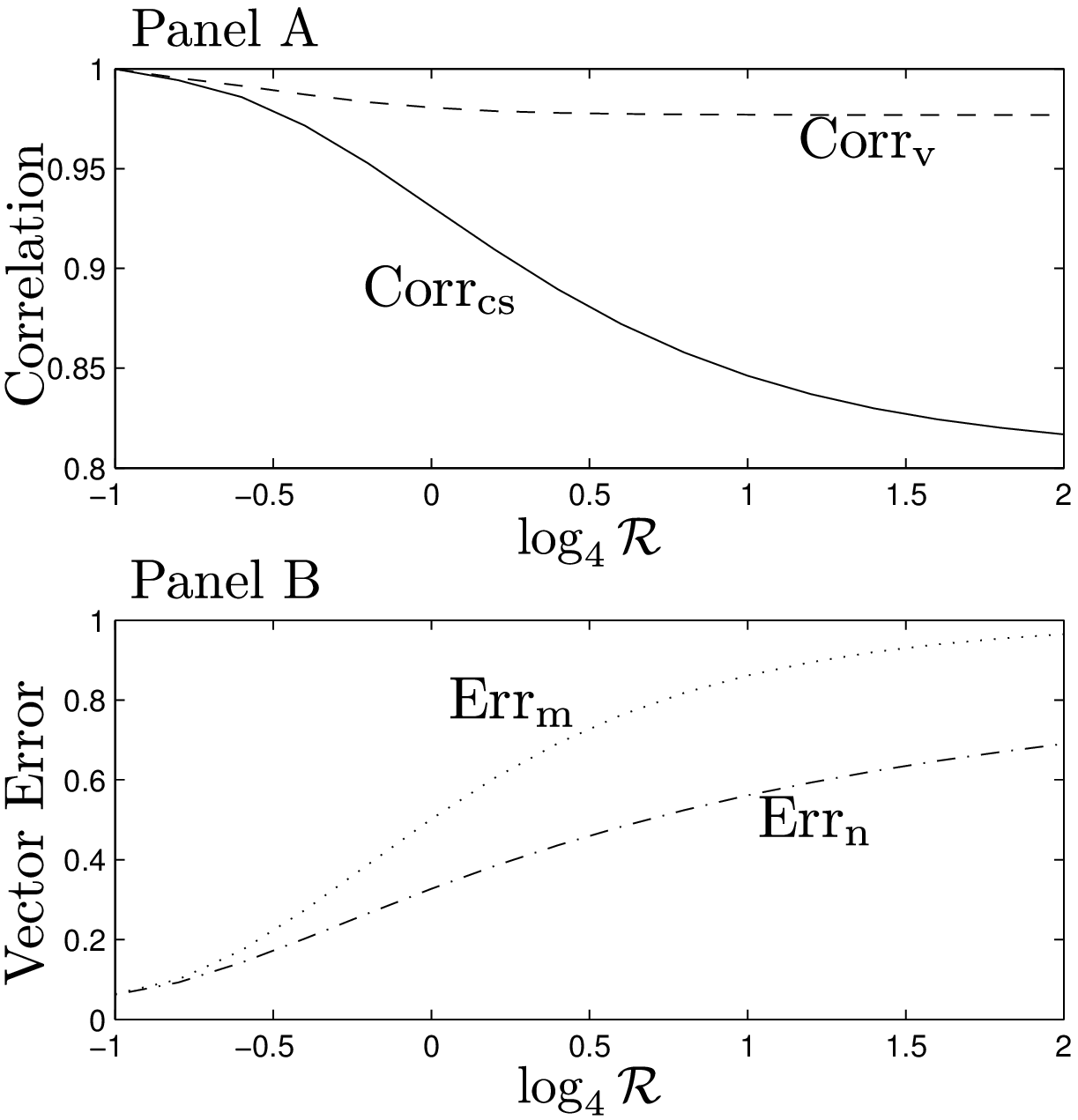}
  \caption{
    Display of four ``diagnostic functions'' as defined by
      equations \eqref{eq:vec-corr} to \eqref{eq:mean-vec-err}
      versus the outer radius $\mathcal{R}$ of the integration
      spatial domain $V$.
  %WLL      {\bf Standard label of abscissa?}
    In Panel A the correlation function for vector magnetic
      fields $\text{Corr}_\text{v}$
      is shown in dashed curve, while the Cauchy-Schwartz
      correlation for vector magnetic fields
      $\text{Corr}_\text{cs}$ is presented by solid curve.
  %WLL      {\bf Seems more sensitive?}.
    In Panel B the normalized error $\text{Err}_\text{n}$
      for magnetic field is by dash-dotted curve and the
      mean error $\text{Err}_\text{n}$ for magnetic field
      is by dotted curve.
    These four ``diagnostic functions'' quantitatively
      show the difference from the semi-analytic one
      when identical boundary conditions and PDEs are
      implemented in the numerical extrapolation.
  }
  \label{fig:diagnostic-force-free}
\end{figure}

We take the spatial integration volume $V$ in eqs. \eqref{eq:vec-corr}
  to \eqref{eq:mean-vec-err}
  %\eqref{eq:norm-vec-err}
  to be a spherical shell whose inner
  radius is $r = r_0=1/\eta_0$ and outer radius is $r=\mathcal{R}$
  which is adjustable.
%   , where we take $\mathcal{R}$ to be variable.
In Figure \ref{fig:diagnostic-force-free}, we illustrate these
  four ``diagnostic functions'' as functions of $\mathcal{R}$.
One can see clearly from Figure \ref{fig:diagnostic-force-free}
  that the correlation functions, especially the Cauchy-Schwartz
  correlation, drops significantly below $1$ as the integration
  volume becomes larger, which reveals that the novel numerical
  solutions deviate from the corresponding semi-analytic ones
  more severely in a larger spatial volume.
%  further areas from center.
The increase of the two error functions for vector magnetic
  fields shown in Panel B reveals a similar trend.

Calculations of these four ``diagnostic functions'' are frequently
  performed in the literature concerning the numerical extrapolation
  scheme of force-free magnetic fields, e.g. \citet{2007SoPh..245..263V},
  \citet{2008SoPh..247...87H} and \citet{2009SoPh..260..109L}.
Although the verification of numerical extrapolation algorithms
%  always
  seem to have good consistency with the results of \citet
  {1990ApJ...352..343L}, our numerical calculations here, taken with
  respect to the novel numerical solution and the semi-analytic
  solution, suggest that significant deviations may also occur
  due to the nonlinearity of the force-free equation.
Possibly those discussions could be responsible for ``not-so-good''
  results of extrapolation tests [e.g. \citet{2008JGRA..11305S90H}
  and \citet{2011SoPh..269...41L}], aiming at reproducing
  separable semi-analytic results in \citet{1990ApJ...352..343L}.
%}

%{\bf The discussion here can be enriched further. Again
% March 30, 2013 Wang Lile at Shanghai}

\section{\ Derivation of the key scalar
  nonlinear elliptical PDE}
\label{sec:scalar-pde-derivation}

%{\bf Please indicate clearly physical
% meanings of derived results in the main
% text and here whenever possible.
%That would help in many ways to grasp
% the essence.  April 6, 2013.}

In this Appendix C, we summarize the major steps
  that lead to the key scalar nonlinear elliptical
  PDE \eqref{eq:u-pde-dimensional}
  \citep[see also][]{2001GApFD..94..249T}.

%First we note that,
In nonlinear PDE system \eqref{eq:vect-quant},
%we adopt a conclusion that
  a divergence-free vector field $\mathbf{A}$ of
  axisymmetry can be generally expressed in terms of
  a toroidal component $A_t$ and a poloidal component
  $A_p$ \citep[e.g.][]{1956PNAS...42....1C} in the
  form of
  \begin{equation}
    \label{eq:div-0-vec-express}
    \mathbf{A} = A_t \nabla \phi + \nabla \phi \times \nabla A_p\ ,
  \end{equation}
  where $A_t$ and $A_p$ are two scalar functions of $r$ and $\theta$,
  and $\nabla \phi = \hat\phivec / (r \sin\theta)$ in
  the spherical polar coordinates $(r,\ \theta,\ \phi)$.
An axisymmetric scalar field ${\cal A}$ with the
  symmetry axis along $\theta=0$ always satisfies the
  condition of $\nabla {\cal A} \cdot \nabla \phi = 0$.
%  {\bf Be careful here. A change of notation! Confirmed
%   with Wang LiLe Mar 30, 2013 at Shanghai}

By projecting the Ohm's law of infinite conductivity,
  i.e. $\evec + \vvec \times \bvec / c = 0$ in the
  azimuthal $\nabla \phi$ direction and assuming
  axisymmetry, we have
  \begin{equation}
    \label{eq:f-function-first-integral}
    \begin{split}
      0 & = \dfrac{(4\pi)^{1/2}}{c} \left( \dfrac{\nabla \phi}{\rho}
      \right) \cdot (\nabla\phi \times \nabla F) \times (\nabla\phi
      \times \nabla \psi)
      \\
      & = -\dfrac{(4\pi)^{1/2}}{c} \left( \dfrac{|\nabla \phi|^2}{\rho}
      \right) \nabla \phi \cdot (\nabla \psi \times \nabla F)\
    \end{split}
  \end{equation}
after straightforward manipulations of vector identities.
%  {\bf Missed a minus sign? Not essential though.
%       Confirmed with LiLe!}
As the only non-zero component of $\nabla\psi\times\nabla F$
  is the $\hat\phivec$ component, we simply have
  $\nabla \psi \times \nabla F=0$ from the last factor
  $\nabla \phi \cdot (\nabla \psi \times \nabla F)=0$
  in condition (\ref{eq:f-function-first-integral}) above.
Similarly, we also derive $\nabla \psi \times \nabla\Phi = 0$
  by projecting the Ohm's law of infinite conductivity along
  the magnetic field $\bvec$ direction.
% {\bf There are three component directions!
%  Which component?}
We readily have $F=F(\psi)$ by integrating
  $\nabla\psi\times \nabla F=0$ and it also follows
  similarly that $\Phi = \Phi(\psi)$.

Another useful relation can be derived by projecting the
  Ohm's law of infinite conductivity along $\nabla\psi$,
  yielding
  \begin{equation}
    \label{eq:i-f-theta-phi-relation-derivation}
    \nabla \psi \cdot \left[ \dfrac{c \nabla \Phi}{(4\pi)^{1/2}} \right]
    =\dfrac{|\nabla \phi|^2}{\rho} (I \nabla \psi\cdot\nabla F - \Theta
    |\nabla \psi|^2 ) \
  \end{equation}
  after straightforward manipulations.
This gives rise to the second line in equation
  \eqref{eq:scalar-relation} by simply inserting
  the two relations $\nabla \Phi = \Phi' \nabla\psi$
  and $\nabla F = F'\nabla\psi$ derived above.

Before more integral relations,
% we obtain more first integrals,
  we invoke below two useful equations.
The first one is the familiar vector identity
  \begin{equation}
    \label{eq:v-dot-grad-v}
    (\vvec \cdot \nabla) \vvec \equiv \dfrac{1}{2} \nabla v^2
    - \vvec \times (\curl \vvec)\ .
  \end{equation}
The second one involves the axisymmetric function $\psi$,
  \begin{equation}
    \label{eq:curl-grad-phi-cross-grad-psi}
    \curl (\nabla \phi \times \nabla \psi) = \Delta^* \psi \nabla \phi\ ,
  \end{equation}
  where $\Delta^* \psi = \div( |\nabla \phi|^2 \nabla \psi )
  / |\nabla \phi|^2$.

Two identities \eqref{eq:v-dot-grad-v} and
  \eqref{eq:curl-grad-phi-cross-grad-psi} are inserted into the
  second line of \eqref{eq:mhd-equ}, i.e. the momentum equation.
Using equations \eqref{eq:vect-quant}, the left-hand side (LHS)
  of equation \eqref{eq:v-dot-grad-v} can be further expanded
  in the form of
  \begin{equation}
    \label{eq:v-cross-curl-v}
    \begin{split}
      & \vvec \times (\curl \vvec) = \dfrac{|\nabla \phi|^2}{2} \nabla
      \left( \dfrac{\Theta}{\rho} \right)^2
      \\
      &  + \left[ \dfrac{F'|\nabla \psi|^2}{\rho} \left( \nabla
          \dfrac{F'}{\rho} \cdot \nabla \psi \right) + \Delta^*\psi
        \left( \dfrac{F'}{\rho} \right)^2 |\nabla \phi|^2 \right] \nabla
      \psi
      \\
      &  - \dfrac{F'}{\rho} \left[ \nabla\dfrac{\Theta}{\rho} \cdot
        (\nabla \phi \times \nabla \psi) \right] \nabla\phi \ .
    \end{split}
  \end{equation}
The specific Lorentz force term $(\curl\bvec)\times \bvec/(4\pi)$
  can also be expanded in the form of
  \begin{equation}
    \label{eq:lorentz-force-expand}
    \begin{split}
      & \dfrac{1}{4\pi} (\curl\bvec) \times\bvec = [(\nabla \phi \times
      \nabla \psi) \cdot \nabla I] \nabla \phi
      \\
      & \qquad\quad - |\nabla \phi|^2 \left( \Delta^* \psi \nabla \psi
       + \dfrac{\nabla I^2}{2} \right)\ .
    \end{split}
  \end{equation}
We first project the momentum equation along the
  azimuthal $\nabla \phi$ direction to obtain
%[equation \eqref{eq:i-f-theta-phi-relation-derivation} is also  utilized]
  \begin{equation}
    \label{eq:force-balance-grad-phi}
    \begin{split}
      0 & = (\nabla \phi \times \nabla \psi) \cdot \bigg\{ \nabla \left[
        \dfrac{I(F')^2}{\rho} + \dfrac{c\Phi'}{(4\pi)^{1/2}}
        F'r^2\sin^2\theta + I \right]
      \\
      & \qquad\qquad -\left[ \dfrac{IF'}{\rho}
        + \dfrac{c\Phi'}{(4\pi)^{1/2}} R^2
        \right] \nabla F \bigg\}\ ,
    \end{split}
  \end{equation}
  where equation \eqref{eq:i-f-theta-phi-relation-derivation} is  used.
We then use the definition of $X$ in the third line of equation
  \eqref{eq:scalar-relation} and $\nabla F \cdot (\nabla \phi
  \times \nabla \psi) = 0$,
  which naturally leads to $\nabla \phi \cdot
  (\nabla \psi \times \nabla X) = 0$ and hence $X=X(\psi)$
  turns out to be another free functional in our MHD formalism.

The momentum equation along the $\bvec$
%  {\bf Three component directions! Which one? }
  direction yields
  \begin{equation}
    \label{eq:force-balance-bvec}
    0 = (\nabla \phi \times \nabla \psi) \cdot \left\{ \nabla \left[
        \dfrac{v^2}{2} + \dfrac{c \Phi'\Theta}{(4\pi)^{1/2}\rho} \right] +
      \dfrac{\nabla P}{\rho} + \nabla \Omega \right\}\ ,
  \end{equation}
where $\nabla \Phi' \cdot (\nabla \phi \times \nabla \psi) = 0$
  because $\Phi' = \Phi'(\psi)$ is a full function of $\psi$ and
  $\Omega = -G\mathcal{M}/r$ is the gravitational potential.
We now impose the condition of incompressibility
  $\div \vvec = 0$ leading to $\rho = \rho(\psi)$
  and thus $\nabla \rho \cdot
  (\nabla \phi \times \nabla \psi) = 0$.
Equation \eqref{eq:force-balance-bvec}
  can be cast into the form of
  \begin{equation}
    \label{eq:force-balance-bvec-integrated}
    0  = (\nabla \phi \times \nabla \psi) \cdot \nabla \left\{
      \dfrac{v^2}{2} + \dfrac{(c \Phi' r \sin\theta)^2}{4\pi[1 - (F')^2/\rho]}
      + \dfrac{P}{\rho} + \Omega \right\}\ ,
  \end{equation}
  where we have invoked the last two lines in equations
  \eqref{eq:scalar-relation} to eliminate $\Theta$.
The following equation is also applied in deriving equation
  \eqref{eq:force-balance-bvec-integrated}, viz.
  \begin{equation}
    \label{eq:force-balance-bvec-integrated-supplement}
    0 = (\nabla \phi \times \nabla \psi) \cdot \nabla \left[
      \dfrac{XF'/\rho}{1 - (F')^2/\rho} \right]\ ,
  \end{equation}
 where $X$, $F'$ and $\rho$ are all full functionals of $\psi$.
If we define the sum of all terms within the curly brace of equation
  \eqref{eq:force-balance-bvec-integrated} to be $P_s/\rho$, we
  thus infer $P_s=P_s(\psi)$ similar to what we did for $F=F(\psi)$
  and $\Phi=\Phi(\psi)$, where the definition of variable $P_s$ is
  consistent with the first line of equation \eqref{eq:scalar-relation}.

The scalar nonlinear elliptical PDE can be derived by projecting
  the momentum equation along $\nabla \psi$ direction.
It is then straightforward yet tedious to obtain
  \begin{equation}
    \label{eq:force-balance-grad-psi}
    \begin{split}
      0 & = \left\{ \nabla \cdot \left[ \left(1 - \dfrac{(F')^2}{\rho}
          \right) \dfrac{\nabla \psi}{r^2\sin^2\theta} \right] +
        \dfrac{F'F''|\nabla \psi|^2} {\rho r^2\sin^2\theta} \right\}
      |\nabla\psi|^2
      \\
      & + \left\{ \dfrac{\rho}{2}\left[ \nabla v^2 -
          \dfrac{\nabla(\Theta/\rho)^2} {r^2\sin^2\theta} \right]
        + \dfrac{\nabla I^2}{2r^2\sin^2\theta} + \nabla P + \rho \nabla
        \Omega \right\}\ .
    \end{split}
  \end{equation}
We simplify PDE (\ref{eq:force-balance-grad-psi}) by substituting
  the definition of $P_s$ [i.e. the first line of equation
  \eqref{eq:scalar-relation}] and the definition of poloidal
  Alfv\'enic Mach number $M$ via $M^2 = (F')^2/\rho$ (see
  subsection \ref{sec:reduction-mhd}).
Straightforward manipulations then yield
  \begin{equation}
    \label{eq:force-balance-grad-psi-simplified}
    \begin{split}
      0 & = (1-M^2) \Delta^* \psi - \dfrac{(M^2)'|\nabla \psi|^2}{2}
      + \dfrac{1}{2}\left( \dfrac{X^2}{1-M^2} \right)'
      \\
      & + r^2 \sin^2\theta (P_s' - \Omega \rho') +
      \dfrac{r^4\sin^4\theta}{2} \left[ \dfrac{(c \Phi')^2 \rho}{
          4\pi(1- M^2)} \right]'\ .
    \end{split}
  \end{equation}
We then utilize transformation \eqref{eq:psi-u-transformation}
  to finally arrive at scalar nonlinear elliptical PDE
  \eqref{eq:u-pde-dimensional}.

\section{Functional analysis for FEM software}
\label{sec:functioncal-ana-fem}

Unlike equation \eqref{eq:u-pde-functional},
   \verb|FreeFem++| uses one or two testing
   variation testers in the FEM space.
Therefore, it gives a functional analysis problem, where
   the ``zero point'' of the following functional
   instead of a ``minimum point'' needs to be found:
\begin{equation}
  \label{eq:my-own-freefem-functional}
  \begin{split}
    \delta\mathcal{J} & = \iint\d\mu\d\eta
    \\
    & \times \bigg[ \dfrac{\eta^2}{(1-\mu^2)} \dfrac{\partial
     (\delta w)}{\partial \eta} \dfrac{\partial u} {\partial \eta}
     +\dfrac{\partial (\delta w)}{\partial \mu} \dfrac{\partial u}
    {\partial \mu} - \dfrac{\mathcal{F}\delta w}
    {\eta^2(1-\mu^2)}\bigg]\ .
  \end{split}
\end{equation}
Here $\delta w$ is the variation tester.
In order to find the zero point, another tester $\delta u$
  should be introduced to calculate $\delta^2\mathcal{J}$:
\begin{equation}
  \label{eq:my-own-freefem-functional-variation}
  \begin{split}
    \delta^2 \mathcal{J} & = \iint\d\mu\d\eta \bigg[
    \dfrac{\eta^2}{(1-\mu^2)} \dfrac{\partial (\delta w)}{\partial
      \eta} \dfrac{\partial (\delta u)} {\partial \eta} +
    \dfrac{\partial (\delta w)}{\partial \mu} \dfrac{\partial
      (\delta u)} {\partial \mu}
    \\
    & - \dfrac{\delta w \delta u} {\eta^2(1-\mu^2)} \left(
      \dfrac{\partial \mathcal{F}(u,\eta,\mu)}{\partial u}
    \right) \bigg]\ .
  \end{split}
\end{equation}
Since the PDE is not linear, we shall expect a Newton iteration
  scheme for the final solution to the non-linear problem.
The computational steps during each iteration are taken as follows:
\begin{enumerate}
\item Calculate $\delta\mathcal{J}$ with current $u$,
  presented as a vector.
\item Calculate $\delta^2\mathcal{J}/\delta u$, presented as
  a matrix.
\item Update the value of $u$ by $u\rightarrow u -
  (\delta^2\mathcal{J} / \delta u)^{-1} \delta \mathcal{J}$ .
\end{enumerate}
The detailed code and conditions may vary from a specific problem to
  another, and from a user to another, but the essentials remain the
  same as outlined above.

\end{document}